\documentclass[sn-apa]{sn-jnl}

\usepackage{graphicx}%
\usepackage{multirow}%
\usepackage{amsmath,amssymb,amsfonts}%
\usepackage{amsthm}%
\usepackage{mathrsfs}%
\usepackage[title]{appendix}%
\usepackage{xcolor}%
\usepackage{textcomp}%
\usepackage{manyfoot}%
\usepackage{booktabs}%
\usepackage{algorithm}%
\usepackage{algorithmicx}%
\usepackage{algpseudocode}%
\usepackage{listings}%
\usepackage{tikz}

\usepackage[T1]{fontenc} 
\usepackage{color}
\usepackage{multirow}
\usepackage{paralist}
\usepackage{array}
\usepackage{wrapfig}
\usepackage{enumitem}
\usepackage{amsmath}
\usepackage{soul}
\usepackage{natbib}

\newcolumntype{L}{>{\centering\arraybackslash}m{2.3cm}}
\newcolumntype{T}{>{\centering\arraybackslash}m{1.5cm}}

\newcolumntype{P}{>{\arraybackslash}m{10.0cm}}
\newcolumntype{H}{>{\arraybackslash}m{4cm}}
\newcolumntype{s}{>{\arraybackslash}m{0.8cm}}
\newcolumntype{C}[1]{>{\centering\arraybackslash}p{#1}}

\definecolor{darkgreen}{rgb}{0.0, 0.5, 0.0} 
\newcommand{\paola}[1]{#1}

\newcommand{\minor}[1]{#1}
\newcommand{\tanja}[1]{#1}

\begin{document}

\title[Article Title]{Enhancing Procedural Writing Through Personalized Example Retrieval: \paola{A Case Study on Cooking Recipes}}


\author*[1]{\fnm{Paola} \sur{Mejia-Domenzain} \small{[0000-0003-1242-3134]}}\email{paola.mejia@epfl.ch}
\author[1]{\fnm{Jibril} \sur{Frej} \small{[0009-0009-0631-0636]}}

\author[1]{\fnm{Seyed Parsa} \sur{Neshaei} \small{[0000-0002-4794-395X]}}
\author[1]{\fnm{Luca} \sur{Mouchel}}
\author[1]{\fnm{Tanya} \sur{Nazaretsky} \small{[0000-0003-1343-0627]}}
\author[1]{\fnm{Thiemo} \sur{Wambsganss}}
\author[1]{\fnm{Antoine} \sur{Bosselut} \small{[0000-0001-8968-9649]}}

\author[1]{\fnm{Tanja} \sur{K\"aser}\small{[0000-0003-0672-0415]}}

\affil[1]{\orgname{School of Computer and Communication Sciences, EPFL}, \city{Lausanne}, \country{Switzerland}}




\abstract{Writing high-quality procedural texts is a challenging task for many learners. While example-based learning has shown promise as a feedback approach, a limitation arises when all learners receive the same content without considering their individual input or prior knowledge. Consequently, some learners struggle to grasp or relate to the feedback, finding it redundant and unhelpful. To address this issue, we present \texttt{RELEX}, an adaptive learning system designed to enhance procedural writing through personalized example-based learning. The core of our system is a multi-step example retrieval pipeline that selects a higher quality and contextually relevant example for each learner based on their unique input. We instantiate our system in the domain of cooking recipes. Specifically, we leverage a fine-tuned Large Language Model to predict the quality score of the learner's cooking recipe. Using this score, we retrieve recipes with higher quality from a vast database of over $180,000$ recipes. Next, we apply \texttt{BM25} to select the semantically most similar recipe in real-time. Finally, we use domain knowledge and regular expressions to enrich the selected example recipe with personalized instructional explanations. We evaluate \texttt{RELEX} in a $2 x 2$ controlled study (personalized vs. non-personalized examples, reflective prompts vs. none) with $200$ participants. Our results show that providing tailored examples contributes to better writing performance and user experience.}




\maketitle

\section{Introduction}
\label{sec:introduction}
Writing, decomposing, and revising texts are critical skills in many daily domains and professional environments. Procedural writing is a form of expository writing that promotes the replicability of procedures and the transfer of knowledge \citep{ambarwati2021procedural}. Procedural texts are ubiquitous in many professions, examples include instruction manuals, algorithmic code~\citep{ambarwati2021procedural}, lab protocols, and cooking recipes~\citep{alviana2019effect}. Unfortunately, many learners struggle to write complete and high-quality procedural texts~\citep{chefs,ambarwati2021procedural}.

Procedural writing is a so-called heuristic domain~\citep{renkl09}, requiring a combination of knowledge of the learning domain (e.g., how to structure a procedural text) and the application domain (e.g., chemistry in the case of lab protocols). This domain dependence prevents the development of a single algorithmic solution for writing good procedural texts. In this context, learners can benefit from learning from examples. Learning from examples enables learners to "borrow" knowledge from others \citep{sweller94} and abstract general rules that can be used to solve similar problems in the future. Prior research has mainly focused on example-based learning applied to highly structured tasks like mathematics and physics \citep{sweller94, HILBERT200854, VANGOG2008211}. Nevertheless, example-based learning has been studied in heuristic domains with no single correct solution \citep{renkl09}. In these contexts, the examples are often enriched to include instructional explanations that can reduce the cognitive load by emphasizing relevant characteristics~\citep{schworm2007learning,VANGOG2008211}. However, the provided examples and instructional explanations are commonly \textit{static} \citep{RENKL2002529}: all learners are provided with the exact same content (e.g., a worked-example by an expert with instructional explanation), independent of their actual skill level. Hence, the provided examples and instructions might be too complex or not relevant to the user, hindering learning and motivation \citep{VANGOG2008211, alamri2020}.

Providing tailored examples and feedback timely, therefore, has the potential to increase learner performance and experience. While there exists a large body of research on optimal task selection in structured domains (e.g., \citet{DBLP:conf/chi/BassenBSTPZGFM20}), only a few works have focused on retrieving examples tailored to the user's context in heuristic domains. Existing research has, for example, employed feature-based similarity metrics \citep{DBLP:journals/nrhm/HosseiniB17, DBLP:journals/tlt/Pelanek20} or unsupervised semantic sentence similarity methods \citep{DBLP:conf/sigir/ZlabingerSHSH20} to retrieve similar educational items. However, the majority of these works focused on retrieving \textit{similar} (in terms of the input text provided by the user) expert-created examples, disregarding the actual skill level of the user.

Furthermore, there is also a vast research on providing personalized explanations and instructions for various writing tasks. Existing tools visualize the revision history of the user's text~\citep{afrin21} or use an underlying domain-specific structure to enrich the user's text with feedback and explanations~\citep{DBLP:conf/chi/WangANCG20}. However, they do not provide suggestions or \textit{examples} on how to correct the shortcomings in the user's text. 

In this paper, we present \texttt{RELEX} (REcipe Learning through EXamples), an effective and scalable learning system for procedural writing using \textit{personalized} example-based learning. \paola{We have instantiated \texttt{RELEX} in the domain of cooking recipes because of its familiarity and practical relevance to culinary students and chef apprentices, as identified by prior work \citep{chefs}.} \texttt{RELEX} features a multi-step pipeline retrieving an example that is 1) relevant for the learner (i.e., similar in terms of topic), 2) of better quality than the learner’s text (i.e., tailored to the learner's skill level), and 3) annotated with explanations and suggestions that the learner’s text is lacking. Our pipeline takes as input the learner's recipe and predicts its quality using a fine-tuned Large Language Model (LLM). Then, it retrieves a set of texts with a higher quality (than the predicted quality) from a database containing over $180'000$ rated recipes. Finally, the most semantically similar recipe is extracted from the retrieved candidate set using \texttt{BM25}.

To evaluate \texttt{RELEX}, we conduct a $2\times2$ controlled study with $200$ participants, in which we manipulate a) the adaptiveness of the provided example and annotations (adaptive vs. non-adaptive example and feedback), and b) the prompts for reflection (reflective prompts vs. none). We also run the same task with a control group receiving static procedural writing support only. With our analyses, we aim to address the following three research questions:  What are the effects of providing a \textit{personalized} example along with adaptive feedback and reflective guidance on learners' \textbf{experience} (RQ1), \textbf{writing performance} (RQ2) and \textbf{revising behavior} (RQ3)?

\noindent Our results indicate that participants who received tailored examples revised their cooking recipes more, wrote them with higher quality, and had a more positive perception of the tool than the users without adaptive feedback.

\section{Related Work and Conceptual Background}
\label{sec:related-work}
In this paper, we present the design and evaluation of a learning system for personalized example-based learning at scale, which is instantiated in the domain of procedural writing. Our study has therefore been influenced by related work in the areas of (1) learning procedural writing skills, (2) example-based learning in heuristic domains, and (3) adaptive learning.

\subsection{Learning Procedural Writing Skills}
\label{sec:procwriting}
Procedural writing, a form of expository writing, facilitates the transfer of knowledge and the replicability of procedures \citep{ambarwati2021procedural}. This type of writing finds its applications in various fields, ranging from life sciences lab protocols to technical documentation and culinary recipes \citep{wieringa1991procedure, chefs, alviana2019effect}.




While procedural writing is highly dependent on the subject matter, previous research ~\citep{wieringa1991procedure, sato2006, Traga_Philippakos_2019, Adoniou_2013} has identified three main qualities of high-quality procedural texts: structure, clarity, and specificity. Structure refers to the organization of the text like having appropriate sections. Clarity involves providing necessary details, and specificity refers to the use of appropriate, domain-specific vocabulary. 

Previous research has found that learners often encounter difficulties when attempting to compose comprehensive and high-quality procedural texts ~\citep{chefs,ambarwati2021procedural}. Common mistakes, in the case of computer documentation and nuclear power plants procedures, are the incorrect order of steps, missing elements, lack of details, or ambiguous words that lead to confusion \citep{wieringa1991procedure}. Similarly, the recipes documented by chef apprentices are often missing ingredients and exhibit a lack of detail and use of specific vocabulary~\citep{chefs}.

Given these challenges in writing procedural texts, the question arises: How can we effectively teach and instruct this skill? Effective feedback mechanisms for procedural writing have received limited attention. One notable investigated mechanism involved feedback through simulation: students were prompted to compose a procedural text detailing how to draw a geometrical figure and subsequently received feedback in the form of the figure drawn based on their instructions \citep{sato2006}.

While there are general learning objectives (structure, clarity, and specificity), the dependence on the domain prevents the development of a single algorithmic solution for writing a good procedural text. In this context, learners can benefit from learning from examples. Previous research has investigated the efficacy of model-based instruction, where students observe a teacher demonstrating and verbally describing the procedure in action. Notably, studies have applied this approach in various scenarios, such as making a peanut butter and jelly sandwich \citep{Traga_Philippakos_2019} and preparing a chicken sandwich \citep{alviana2019effect}. Encouragingly, both works reported positive effects on the quality of procedural writing resulting from the implementation of the demonstration technique. Surprisingly, despite the proven benefits of using written worked examples in other genres, such as argumentation skills \citep{schworm2007learning}, their potential application in procedural writing remains largely unexplored.

\subsection{Example-Based Learning in Heuristic Domains}
\label{sec:examplebased}
Example-based learning is an effective method to acquire knowledge by observing and/or imitating what other people do, say, or write~\citep{sweller94}. It allows learners to build a cognitive schema of how problems should be solved. In addition, learners can abstract general rules from the examples and ultimately transfer and adapt them to other problems~\citep{vangog2010}.  The vast majority of research on example-based learning has studied their effectiveness in well-structured tasks, such as algebra~\citep{sweller94} and physics~\citep{VANGOG2008211}. More recently, worked-examples and solved-examples have been applied to non-algorithmic learning domains such as argumentative writing~\citep{schworm2007learning} and mathematical proof finding~\citep{HILBERT200854}. In heuristic domains~\citep{renkl09}, where no algorithmic solution can be provided (e.g. cooking recipes), learners acquire heuristics that help them find a solution. Examples in heuristic domains require learners to process two different content levels: (1) the learning domain (i.e., how to structure the solution) and (2) the exemplifying domain (i.e., the topic). In the case of cooking recipes, learners need to understand how to structure a procedural text (learning domain: procedural writing) and be familiar with the cooking domain (the exemplifying domain). Given the two content levels, these examples are referred to as \textit{double-content}. In structural domains, worked-examples are usually annotated with the steps to solve the problem. In contrast, the \textit{double-content} examples tend to be \textit{enriched} with self-explanation prompts and/or additional instructional explanations.\vspace{1mm}

\noindent \textbf{Reflective Prompts}.
According to the \textit{self-explanation effect}, learners benefit more from the examples if they can actively explain the examples to themselves~\citep{wong2002effects}. Furthermore, the quality of the self-explanations determines what is learned from the examples \citep{CHI1989145}. However, frequently, learners' self-explanations are superficial or passive. Thus, the application of prompts is a possible intervention to increase the quality and depth of the explanations. These prompts should stimulate the active processing of learning materials and direct attention to the central aspects~\citep{schworm2007learning}. The use of self-regulated learning (SRL) prompts has been shown to foster conceptual knowledge \citep{Roelle2012}. Furthermore, SRL prompts (i.e., \textit{which aspects of the learning materials do you find interesting, useful, and convincing, and which not?}) have been used to help the learner focus on the central elements of examples~\citep{nuckles09} or to guide learners to diagnose their deficiencies and be critical~\citep{Fan_Luo_Menekse_Litman_Wang_2017}.\vspace{1mm}

\noindent \textbf{Instructional Explanations}. Instructional explanations are another possibility to enrich examples. It has been demonstrated that in a first learning phase, instructional explanations improve the learning outcomes compared to when there are no explanations provided ~\citep{VANGOG2008211}. However, these explanations can be detrimental later in the learning, since the provided information soon becomes redundant and the explanations increase the cognitive load and hinder learning. Instructional explanations have the following disadvantages in comparison to self-explanation~\citep{RENKL2002529}: (1) they are not adapted to the learner's prior knowledge, so they can be redundant or too complex and hard to understand; (2) they are often not timely and therefore hard to integrate as part of the ongoing learner's activities.

In a $2\times2$ study on the effect of self-explanation prompts and instructional explanations, the group that received only self-explanation prompts had the most favorable learning outcomes, whereas the group that received instructional explanations had the highest perception of learning~\citep{schworm2007learning}. Nevertheless, the authors did not examine the use of adaptive instructional explanations.  A first step in this direction has been taken by providing so-called faded examples in geometry learning~\citep{Schwonke_Renkl_Krieg_Wittwer_Aleven_Salden_2009}. Students were shown complete worked-out examples at first; over time steps from the example were gradually removed. However, the missing steps and the selected examples were pre-determined and not chosen adaptively depending on the students.

To summarize, the provided examples, the reflective prompts, and the instructional explanations are commonly static: all learners are provided with the same content (e.g., a worked-example by an expert with instructional explanation). The examples and explanations are (1) not adapted to the learner's prior knowledge, so they can be redundant and hence hinder learning \citep{VANGOG2008211} and (2) not timely and relevant, hence decreasing engagement \citep{alamri2020}. Providing \textit{personalized examples and instruction} in a timely manner therefore has the potential to improve learning.

 \subsection{Adaptive Learning}
 \label{sec:perslearning}
Providing personalized examples and adaptive annotations and explanations translates into providing 1) personalized content (the example) and 2) personalized instruction.\vspace{1mm}

\noindent \textbf{Personalized Content}.
In content level adaptation, the learning objects (e.g., examples, tasks) are selected and adapted based on the content (e.g., current task, answer, knowledge state) of the user~\citep{Premlatha_Geetha_2015}. One approach to providing personalized content is to \textit{retrieve} a tailored example from an existing collection. The collection consists of all the examples available, the query is the user's context and the system ranks examples in the collection based on their similarity with the user's context. Depending on the task to be learned, the user's context can be the current task, the answer, the learner's knowledge or any combination of these. Example retrieval involves three steps: \textbf{(1)} computing a similarity between the learner's context and examples from the collection, \textbf{(2)} ranking the examples based on their similarity and \textbf{(3)} presenting the most similar or top-$k$ examples to the learner. For instance, \citet{DBLP:journals/nrhm/HosseiniB17} used semantic-level similarity-based linking to recommend personalized examples to programming learners. ~\citet{DBLP:journals/tlt/Pelanek20} explored feature-based (such as the occurrence of domain-specific keywords) and performance-based measures to compare the similarity of educational items in various domains. Furthermore, ~\citet{DBLP:conf/sigir/ZlabingerSHSH20} provided crowdworkers with personalized examples: they used unsupervised semantic sentence similarity methods to retrieve tailored expert-labeled examples.

Obtaining high-quality expert examples for learning purposes can be challenging and costly. In such cases, peer examples serve as an alternative, which, despite their potential loss in quality, can prove more effective in a learning scenario \citep{DBLP:conf/chi/DoroudiKBH16}. However, evaluating the quality of peer examples poses its own challenge, as the perception of good quality varies among raters, tasks, and genres \citep{Wilson_Olinghouse_Andrada_2014}. To address this issue, recent research has explored the application of LLMs, like BERT \citep{DBLP:conf/naacl/DevlinCLT19} or GPT-models \citep{DBLP:conf/nips/BrownMRSKDNSSAA20}, for tasks such as automatically scoring essays \citep{Mayfield_Black_2020}, rating recipe nutritional quality \citep{HU2022}, and evaluating text generation \citep{DBLP:conf/acl/SellamDP20}. These LLMs, being at the forefront of natural language processing (NLP) tasks \citep{DBLP:conf/naacl/DevlinCLT19, liu2019roberta, DBLP:conf/nips/BrownMRSKDNSSAA20}, offer a promising approach to predict the quality of examples in heuristic domains.

\noindent \textbf{Personalized Instruction}.
In contrast to generic instruction, personalized instruction (or feedback or explanation) is dynamic, which means that different learners will receive different information~\citep{Bimba_Idris_Al-Hunaiyyan_Mahmud_Shuib_2017}. While there is a range of research on providing personalized feedback and hints in structured domains such as mathematics \citep{Paassen_Hammer_Price_Barnes_Gross_Pinkwart_2018} or programming \citep{Ahmed_Srivastava_Sindhgatta_Karkare_2020}, less work has focused on giving automated fine-grained suggestions and explanations in heuristic domains such as expository writing.

Existing NLP-based writing support tools often provide holistic feedback on higher-level properties of the text such as grammar errors, fluency, or coherence (e.g., \textit{Grammarly}~\citep{grammarly}). To provide more detailed guidance, other tools adopt alternative approaches. For instance, ArgRewrite \citep{afrin21} visualizes revision histories by annotating a side-by-side comparison of two drafts, providing revision suggestions at the sentence and sub-sentence level. In contrast, ArguLens \citep{DBLP:conf/chi/WangANCG20} utilizes a domain-specific structure by imposing an argumentation-enhanced representation, breaking the user text into argumentation components and standpoints. Despite these valuable contributions, none of the existing systems combine adaptive instruction with a comparison example that could leverage the potential of example-based learning.

\section{RELEX - Learning With Personalized Examples}
\label{sec:design}
\begin{figure}[t]
    \centering
       \includegraphics[clip, trim=0.0cm 0.0cm 13.9cm 0.0cm, width=1.00\textwidth]{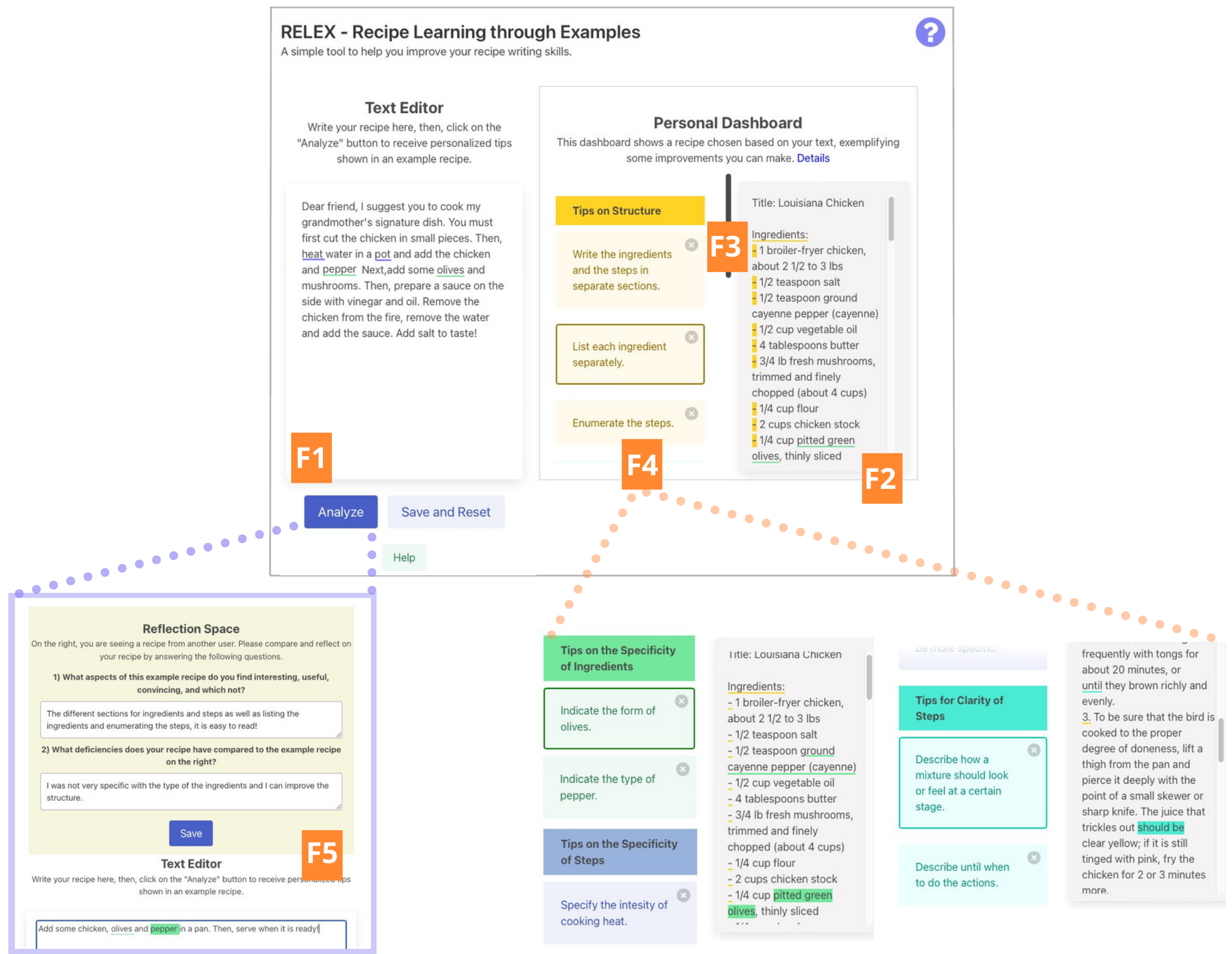}
 \caption{\paola{Interaction flow:} A learner requests feedback (F1) and receives a tailored example recipe (F2) with highlighted in-text elements (F3) and personalized explanations (F4). The learner is also prompted to reflect on the strengths and weaknesses of the recipes (F5).}
     \label{fig:functionalities}
\end{figure}

To study the effect of \textit{personalized} example-based feedback on learners\paola{'} writing performance, revision behavior, and learning experience, we designed \texttt{RELEX} (REcipe Learning through EXamples). The primary purpose of \texttt{RELEX} is to facilitate procedural writing by providing students with tailored examples, accompanied by relevant annotations and reflective prompts. The tool aims to address three key aspects of procedural writing: (a) the organization and structure of texts, (b) the provision of requisite details for enhanced clarity, and (c) the appropriate utilization of specific vocabulary. In the following, we will describe the two main components of \texttt{RELEX}, the user interface and the personalized example retrieval pipeline.

\subsection{User Interface}
The user interface of \texttt{RELEX} is illustrated in Figure \ref{fig:functionalities}\footnote{\paola{The demo version of \texttt{RELEX} is available at \url{https://go.epfl.ch/relex}}}. \paola{The main interface is shown on the upper part of the figure. The interface is split into two main panels: the Text Editor (left) where learners can write or edit recipes and request feedback by clicking "Analyze" ($F_1$) and the Personal Dashboard (right) displaying a selected example recipe with personalized annotations. This Personal Dashboard is again split vertically into two sections, listing suggestions to improve the recipe on the left ($F_4$) and showing the example recipe ($F_2$) with missing aspects in the learner's recipe highlighted ($F_3$) on the right. Below the main interface (Figure \ref{fig:functionalities} bottom right), other types of recipe improvement tips together with a fragment of the example recipe that fulfills these suggestions are shown. More specifically, the bottom middle panel shows examples of tips on the specificity of ingredients and steps; and to the right, there are examples on the clarity of the steps. Finally, the Reflection Space ($F_5$, bottom left) invites the user to carefully study the example recipe with the question \textit{What aspects of this example recipe do you find interesting, useful, convincing, and which not?}; and compare it to their own recipe with the question \textit{What deficiencies does your recipe have compared to the example recipe on the right?}.} The (synthetic) example in Figure \ref{fig:functionalities} illustrates these design functionalities. The learner has asked for feedback ($F_1$) on a recipe including chicken and is provided a similar recipe ("Louisiana Chicken") of higher quality (immediately visible by its clear structure) as an example ($F_2$). The highlights indicate the missing structural elements (for example, "List each ingredient separately", $F_3$) and the left-hand pane of the personal dashboard suggests other tips on the structure like enumerating the steps ($F_4$). Other examples of $F_3$ and $F_4$ are shown below the main interface. In addition, the reflection panel ($F_5$) opens to the bottom left where the user answers the questions.  The five design functionalities of \texttt{RELEX} (see Table \ref{tab:functionalities}) are based on design requirements derived from user interviews as well as from literature.

\vspace{3mm}
\noindent \textbf{User Requirements}. Given that the users should be the main focus of a design effort~\citep{Cooper_Reimann_Cronin_Cooper_2007}, we conducted ten semi-structured user interviews (female-identifying: $6$, male-identifying: $4$) to better understand the specific user needs when using example-based learning in the context of procedural writing\footnote{The detailed interview questions can be found on \url{https://github.com/epfl-ml4ed/relex/blob/main/docs/user-interviews.pdf}}. Participants described their past experiences with writing procedural texts, which included tutorials, lab protocols, technical manuals, and cooking recipes. One common difficulty they encountered when writing procedural texts was being too vague, missing steps, and having the readers struggle to reproduce the instructions they had written. 

From these semi-structured interviews, we derived $22$ user stories. The stories contained a multitude of detailed suggestions, such as the type of colors used for highlighting elements of the text or the request to see explanations for the highlighted elements. We clustered the different user stories based on the underlying topic and obtained five groups, from which we formed the following user requirements:

\begin{itemize}[leftmargin=3.5em]
    \item [($U_1$)] Examples should be relevant and similar so that the users can relate to them.
    \item [($U_2$)] Users should be able to see more than one example in order to generalize and abstract the relevant elements.
    \item [($U_3$)] The important elements of the text should be highlighted with different colors (indicating what each color means) to stimulate active processing. 
    \item [($U_4$)] The mechanism should have interactive explanations (e.g., when the mouse scrolls on top or clicks) of the highlighted text in the form of suggestions or questions (that can be dismissed) to help learners understand the underlying structure of the example.
    \item [($U_5$)] The mechanism should include self-explanation and self-reflection prompts to foster active understanding of the example.
\end{itemize}

\vspace{3mm}
\noindent \textbf{Literature Requirements}. 
After deriving the user-centric requirements, we complemented them with the large body of literature on example-based learning (described in detail in Section \ref{sec:examplebased}). The impact of this approach is highly dependent on the design of the examples utilized. With this regard, previous research examined various design aspects such as self-explanation prompts~\citep{schworm2007learning}, content guidance~\citep{renkl09}, and highlighting~\citep{ringenberg2006scaffolding}. In their review paper,~\citet{vangog2010} synthesized these aspects and provide design guidelines for example processing. Similarly,~\citet{RENKL2002529} derived design principles for instructional explanations. Drawing from the insights of these two review papers, we establish the literature-based design requirements of \texttt{RELEX}:
\begin{itemize}[leftmargin=3.5em]
    \item [($L_1$)] Active processing of examples should be stimulated by emphasizing important aspects of the procedure. This will help learners understand the underlying structure and transfer that knowledge to a different task~\citep{vangog2010}.
    \item [($L_2$)] Learners should be instructed to self-explain the example in order to foster active processing and understanding~\citep{vangog2010}.
    \item [($L_3$)] Examples and explanations should be presented on learners' demand to ensure that they are appropriately timed and used in ongoing knowledge-building activities~\citep{RENKL2002529}.
    \item [($L_4$)] Explanations should be short and minimalist to reduce the effort to process them~\citep{RENKL2002529}.
    \item [($L_5$)] Explanations should not tell learners things that they already know or do not need to know~\citep{RENKL2002529}.
\end{itemize}

\begin{table}[t]
    
\centering
\begin{tabular*}{\textwidth}{@{}p{0.7cm}p{8.1cm}C{1.8cm}C{1.5cm}@{}}
\toprule 
\multirow{2}{*}{\textbf{Code}} & \multirow{2}{*}{\textbf{Functionality}}  & \multicolumn{2}{c}{\textbf{Requirements}}   \\
& & \textit{Literature} & \textit{Users}\\
 \midrule 
  \multirow{3}{*}{$F_1$} & The "Analyze" button allows learners to request feedback on demand. In addition, learners can click on it multiple times and see different suitable examples and explanations. & \multirow{3}{*}{$L_3$} & \multirow{3}{*}{$U_2$} \\
  \midrule
\multirow{3}{*}{$F_2$} &
  Learners are shown a better and similar recipe chosen based on their recipe using our personalized example retrieval pipeline (see Section \ref{sec:pipeline}). & & \multirow{3}{*}{$U_1$} \\
   \midrule 
\multirow{3}{*}{$F_3$} &
  The example recipe highlights with separate colors some aspects that the learner's recipe is missing regarding the structure, clarity, and specificity. & \multirow{3}{*}{$L_1$, $L_5$} & \multirow{3}{*}{$U_3$} \\
   \midrule 
\multirow{2}{*}{$F_4$} &
  The in-text highlighted elements have short explanations in the form of suggestions. These tips can be deleted if not relevant. & \multirow{2}{*}{$L_4$} &
  \multirow{2}{*}{$U_3, U_4$}\\
   \midrule 
\multirow{2}{*}{$F_5$} &
  Learners are instructed to self-explain the example and reflect on their recipe deficiencies in the "Reflection Space".  & \multirow{2}{*}{$L_2$} & \multirow{2}{*}{$U_5$}\\
 \bottomrule
\end{tabular*}
 \caption{Overview of design functionalities and associated user and literature requirements.}
    \label{tab:functionalities}
\end{table}

\paola{Table \ref{tab:functionalities} illustrates the relationship between user and literature requirements and the corresponding functionalities of the tool. The design of these functionalities focused on meeting both the needs identified in the relevant literature and those expressed by users, with the goal of creating a tool that is both educationally effective and user-friendly. Specifically, we began by considering user requirements and then incorporated requirements derived from the literature where applicable. For instance, $F_1$ caters to the users' need for accessing multiple examples ($U_2$) and also aligns with the literature's emphasis on the availability of on-demand examples ($L_3$). Similarly, $F_3$ fulfills the requirement of highlighting important aspects of the learning material ($L_1$) by using different colors, a feature specifically requested by users ($U_3$), while also ensuring that redundancies are minimized ($L_5$). Furthermore, $F_4$ supports the users' desire for explanations ($U_4$) and the use of varied colors ($U_3$), while also adhering to the recommendation for brevity and minimalism in explanations ($L_4$). Additionally, $F_5$ addresses the users' preference for self-explanation prompts ($U_5$) in line with literature insights ($L_2$). Finally, $F_2$, which responds to the users' desire for relevant and similar examples, represents an innovative aspect of our work. }

\subsection{Personalized Example Retrieval Pipeline}
\label{sec:pipeline}
To retrieve tailored examples for learners, we propose a multi-step recipe selection pipeline.  Our pipeline retrieves a \textit{personalized} example recipe that satisfies the following constraints: 1) describing a similar dish, thus relevant to the learner, 2) of higher quality than the learner’s recipe, 3) annotated with explanations and suggestions based on identified weaknesses of the learner's recipe, and 4) retrieved in real-time.

\begin{figure}[t]
    \centering
    \includegraphics[clip, trim=0.0cm 3.0cm 0.5cm 0.0cm, width=1\textwidth]{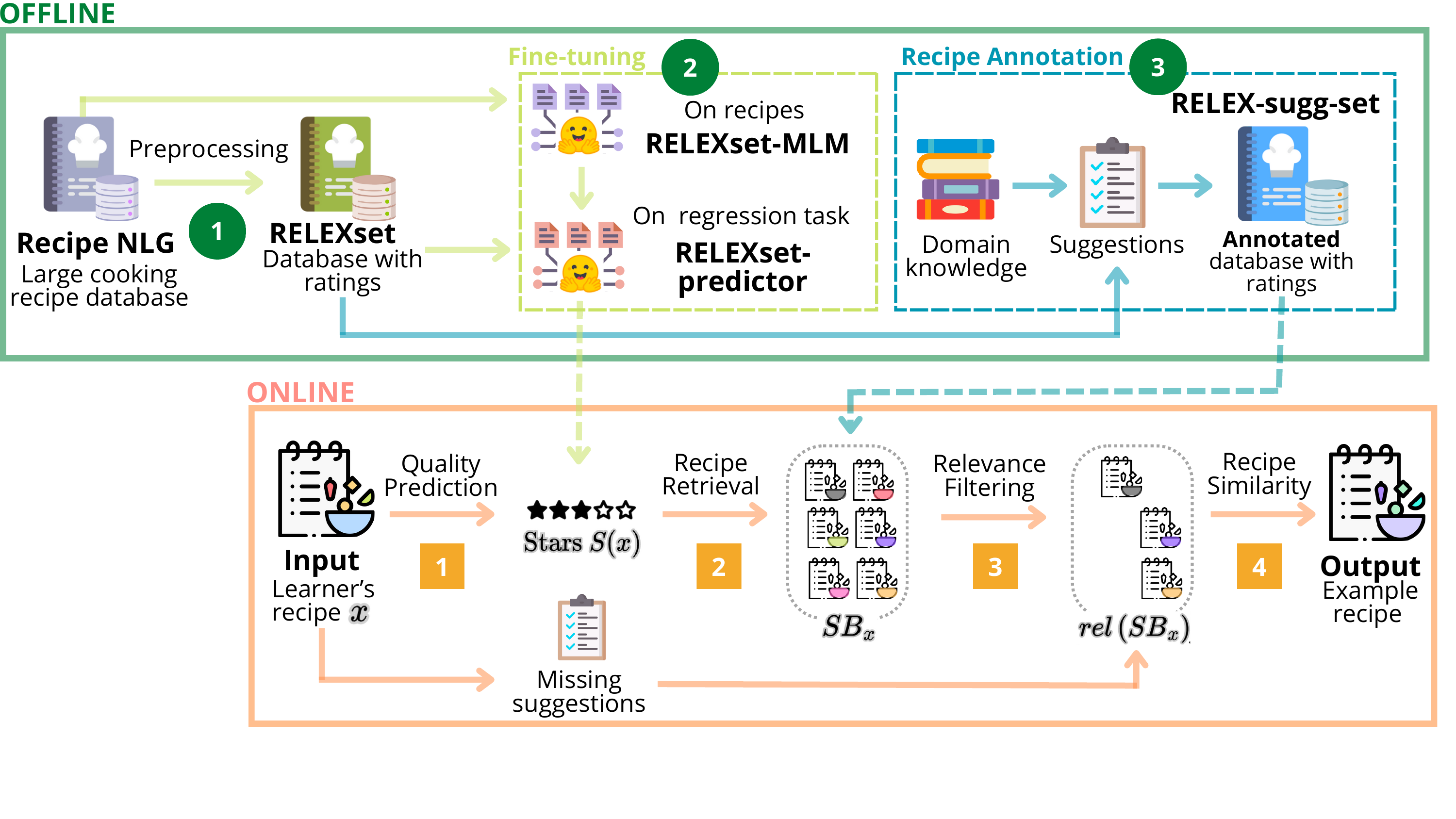}
    \caption{Example selection pipeline. In the offline part, \raisebox{.5pt}{\textcircled{\raisebox{-.9pt} {1}}} cooking recipes are pre-processed to obtain \texttt{RELEXset}; \raisebox{.5pt}{\textcircled{\raisebox{-.9pt} {2}}} \paola{a general domain language model is fine-tuned to get a recipe quality predictor}; \raisebox{.5pt}{\textcircled{\raisebox{-.9pt} {3}}} cooking recipes are annotated using domain knowledge and regular expressions. In the online part, \fbox{1} the quality of the input recipe is predicted, \fbox{2} \paola{recipes of better quality are retrieved}, \fbox{3} \paola{only relevant recipes according to the suggestions are kept}, and \fbox{4} recipe similarity is used to output the most similar example recipe.}
    \label{fig:pipeline}
\end{figure}

\paola{Hence, both the retrieved example and the highlighted suggestions are tailored to the learner’s content (e.g., the type of recipe) and skill level (e.g., the quality of the recipe). The pipeline is illustrated in Figure ~\ref{fig:pipeline}. It features an offline and an online component. The offline component (top, in green) consists of three main steps:
\begin{itemize}
    \item[\raisebox{.5pt}{\textcircled{\raisebox{-.9pt} {1}}}] \textbf{Preprocessing:} a large cooking recipe database (\texttt{RecipeNLG}) is preprocessed to obtain the ratings for each recipe. We denote the resulting dataset of rated recipes as \texttt{RELEXset}.
    \item[\raisebox{.5pt}{\textcircled{\raisebox{-.9pt} {2}}}] \textbf{Fine-Tuning:} a general domain language model is fine-tuned on all recipes from \texttt{RecipeNLG}. This model is further fine-tuned on the regression task to predict the stars from \texttt{RELEXset}. We call this fine-tuned model \texttt{RELEXset-predictor}.
    \item[\raisebox{.5pt}{\textcircled{\raisebox{-.9pt} {3}}}] \textbf{Recipe Annotation:} the recipies from \texttt{RELEXset} are annotated using writing suggestions obtained from experts. We denote the resulting dataset of annotated, rated recipes as \texttt{RELEX-sugg-set}.
\end{itemize} 
The online component (orange, Fig.~\ref{fig:pipeline} bottom), processes the learner's recipe $x$ in four steps:
\vspace{1mm}
\begin{itemize}
    \item[\fbox{1}]\textbf{Quality Prediction:} the stars (quality) of $x$, denoted as $S(x)$, is predicted using \texttt{RELEXset-predictor}.
    \item[\fbox{2}]\textbf{Recipe Retrieval:} all recipes of higher quality than $x$ ($SB_x$) are retrieved from \texttt{RELEX-sugg-set}.
    \item[\fbox{3}]\textbf{Relevance Filtering:} only relevant recipes according to the missing suggestions are kept. We denote this filtered set as $rel(SB_x)$.
    \item[\fbox{4}]\textbf{Recipe Similarity:} recipe similarity is computed to output the most similar example recipe from $rel(SB_x)$. 
\end{itemize}
In the following subsections, we describe each step of the offline and online components in detail.
}

\subsubsection{Offline Training and Annotation}
\label{sec:offlinepipe}
As seen in Fig.\ref{fig:pipeline}, the offline phase consists of three steps: Preprocessing the database, training the cooking domain LLM for rating prediction, \paola{and annotating the recipes with improvement suggestions}.
\paola{Specifically, we quantify the quality of the recipes using crowd-sourced ratings, which allows us to sort the recipes based on the community's perception. Then, we train a model to predict the rating (in the form of stars) a new recipe $x$ would obtain, enabling us to extract recipes of higher quality than the recipe $x$ from the database.} \vspace{1mm}

\noindent \textbf{\paola{Preprocessing}}. We use \texttt{RELEXset}, a database composed of rated cooking recipes\footnote{\texttt{RELEXset} can be downloaded from \minor{https://github.com/epfl-ml4ed/relex/}readme.md}. The recipes were extracted from \texttt{RecipeNLG}~\citep{bien-etal-2020-recipenlg}, a publicly available dataset of clean and formatted versions of cooking recipes. Ratings are real numbers from 0 to 5. They were obtained from \url{food.com}, an online recipes site~\citep{majumder-etal-2019-generating}. We remove from \texttt{RELEXset} all recipes with no ratings. As a result, \texttt{RELEXset} contains more than $180,000$ clean and formatted recipes with more than $700,000$ user ratings. One common problem with user ratings is that different users adopt different criteria and rating scales. Some users might, for example, be more tolerant than others and give higher ratings in general~\citep{jin2004}. To mitigate this bias, following common practices in collaborative filtering models~\citep{jin2004}, we standardize the ratings per user. We denote by $R_y(x)$ the rating of user $y$ for recipe $x$ and by $\hat{R}_y$ the average rating of user $y$ across all recipes. Standardization consists in centering $R_y(x)$ around $\hat{R}_y$ with a unit standard deviation as follows: $\widehat{R}_y\left(x\right) = \left(R_y\left(x\right) - \overline{R}_y\right)\bigg/\sqrt{\sum_{z \in \mathcal{X}} \frac{1}{|\mathcal{X}|} \left( R_y(z) - \overline{R}_y \right)^2}$ with $\mathcal{X}$ the set of all recipes in \texttt{RELEXset}. As the standardized rating cannot be computed when the standard deviation is $0$, users who have only rated one recipe are automatically excluded from the analysis. To obtain a unique rating $S(x)$ associated with each recipe, we average the standardized ratings across all users: $S(x) = \sum_{y \in \mathcal{Y} }\widehat{R}_y\left(x\right)\big/|\mathcal{Y}|$ with $\mathcal{Y}$ the set of all users in \texttt{RELEXset}. In the remaining part of the paper, we use “stars” to refer to the averaged user-standardized ratings.\vspace{1mm}

\noindent \textbf{Fine-Tuning on Recipes}. Given that the recipes consist of text, we follow the recent advances in NLP~\citep{DBLP:conf/naacl/DevlinCLT19, liu2019roberta, Sanh2019DistilBERTAD, DBLP:conf/nips/BrownMRSKDNSSAA20} and use a pre-trained LLM to predict the quality (starts) of a recipe. \paola{The choice of pre-trained LLM is based on performance and efficiency. On the one hand, BERT, a widely-recognized LLM, employs self-attention mechanisms to generate context-aware word representations \citep{DBLP:conf/naacl/DevlinCLT19}. While BERT offers robust performance, RoBERTa, an enhanced version, surpasses it in various NLP benchmarks due to extensive training and hyperparameter optimization \citep{liu2019roberta}. On the other hand, RoBERTa's computational demands are substantial, making it less ideal for real-time applications. To balance performance and efficiency, we opt for DistilRoBERTa, a streamlined version of RoBERTa \citep{Sanh2019DistilBERTAD}. Developed through knowledge distillation, DistilRoBERTa maintains much of RoBERTa's efficacy but with reduced complexity, making it more suitable for our requirement of real-time prediction without relying on GPUs. This is in line with studies suggesting that increased prediction time can negatively impact user experience \citep{DBLP:conf/amcis/Nah03}. Therefore, we initialize our predictor with the \texttt{distilroberta-base} checkpoint from HuggingFace's transformers \citep{DBLP:journals/corr/abs-1910-03771}.}

It is worth noting that, \texttt{distilroberta-base} was trained on general texts from the internet and not specifically in the cooking domain. Following common practices~\citep{DBLP:conf/acl/GururanganMSLBD20,DBLP:conf/cncl/SunQXH19}, before fine-tuning the model for rating recipes, we first adapt \texttt{distilroberta-base} to the cooking domain by fine-tuning it on a Masked Language Modeling (MLM) task on the entire set of recipes from \texttt{RecipeNLG}. We will refer to the resulting model as \texttt{RELEXset-MLM}.\vspace{1mm}

\noindent \textbf{Fine-Tuning on a Regression Task}. \paola{Given that we want to predict the averaged user-standardized rating (stars) of a recipe, we formulate the prediction stage as a regression task: for any given recipe denoted as $x$, the predictive model should output a real-valued star rating symbolized as $S(x)$. }
Thus, we fine-tune \texttt{RELEXset-MLM} to predict the number of stars of recipes in \texttt{RELEXset}. \minor{We will refer to the obtained model as \texttt{RELEXset-Predictor}\footnote{\texttt{RELEXset-MLM} is available at \url{https://huggingface.co/paola-md/RELEXset-MLM} and \texttt{RELEXset-Predictor} is available at \url{https://huggingface.co/paola-md/RELEXset-Predictor/}}}. \minor{The model has six transformer layers, each with a hidden size of 768, and employs 12 attention heads. The intermediate layers in the transformers have a size of 3072. Moreover, the model uses GELU as its activation function and dropout rates for both attention probabilities and hidden layers are set to 0.1}\footnote{The architecture configuration is available at \url{https://huggingface.co/paola-md/RELEXset-Predictor/blob/main/config.json}}. Following, we use a fully connected neural network with one hidden layer that takes the [CLS] token final embedding as input and outputs the number of stars $S(x)$. We optimize both \texttt{RELEXset-MLM} and \texttt{RELEXset-Predictor} using \texttt{adam}~\citep{DBLP:journals/corr/KingmaB14} with early stopping. Both \texttt{RecipeNLG} and \texttt{RELEXset} are split into train/validation/test sets with a ratio of 80/10/10. This ratio was chosen to provide sufficient data for training while also allowing adequate samples for validation and testing. Given the complexity of the model, the 80/10/10 split ensures that more data is available for training. Furthermore, given the large size of the dataset, $10\%$ of the data points used for validation and testing are sufficient to validate and test effectively. \minor{We used the Kolmogorov–Smirnov test\footnote{After a significant Shapiro-Wilk test on the three sets ($p=0$)}, a nonparametric test of the equality of continuous probability distributions, to verify that there were no significant differences (train vs validation: $p=.36$, train vs test: $.91$, validation vs test: $p=.75$) between the label distributions in the train, validation and test sets\footnote{Visual validation available at: https://github.com/epfl-ml4ed/relex/blob/main/docs/split-verification.ipynb}.} Learning rate, batch size, and weight decay were selected on the validation set using grid search from \{1e-6, 1e-5, 2e-5, 3e-5, 5e-5, 1e-4\}, \{32, 64, 128, 256, 512\} and \{0.01, 0.02, 0.03, 0.05, 0.08, 0.1\} respectively. \minor{We chose the best model (hereafter referred as \texttt{RELEXset-Predictor}) based on the validation loss and tested its performance on the hold-out test set.}  \texttt{RELEXset-Predictor} achieved a mean absolute error (MAE) of $0.39$ on the test set, which is \minor{slightly} better than the baseline MAE of $0.42$ (simply predicting the mean). \minor{Despite the difference not being significant, \texttt{RELEXset-Predictor} has the ability to generalize to new, unseen data, making it a more reliable tool for making predictions in real-world scenarios than the static baseline predictor. As outlined in Section 3.2.2, the subsequent stages of the pipeline are designed to address the prediction uncertainties by selecting recipes that fall within a quality range set above the MAE threshold to ensure that the recipe is perceived as better by the users.}


\vspace{1mm}

\noindent \textbf{Recipe Annotation}. After choosing a targeted example recipe, we enrich the recipe with suggestions on how to improve the text. These suggestions are based on the three main aspects of high-quality procedural text \citep{wieringa1991procedure, sato2006, Traga_Philippakos_2019}: structure (i.e., clear organization of the text), clarity (i.e., appropriate degree of detail), and specificity (i.e., proper use of technical terms). The suggestions can be divided into general suggestions concerning the learning domain (i.e., how to write procedural text) and into suggestions specific to the exemplifying domain (cooking recipes). The domain-general suggestions are derived from the main qualities of good procedural text identified in previous work \citep{wieringa1991procedure, sato2006, Traga_Philippakos_2019}. The domain-specific suggestions are derived from "The Recipe Writer's Handbook, Revised and Expanded"~\citep{handbook}. In this handbook, two recipe book editors give punctual recommendations on how to write a good recipe in terms of the learning objectives (structure, specificity, and clarity). We use the keywords ``specify'' and ``indicate'' to retrieve $45$ suggestions from the handbook. Table \ref{tab:annotations} lists all the domain-general suggestions as well as examples of domain-specific suggestions. There are four suggestions related to the structure and three suggestions related to the clarity of the text. For these two categories, there is a direct mapping between domain-general and domain-specific annotations. There are in total $38$ recipe-specific suggestions related to the specificity of the steps and material\footnote{Complete list of suggestions and classification rules available at \minor{https://github.com/epfl-ml4ed/relex/}docs/recipe-suggestions-rules.pdf.}.

\begin{table}[t]
    \centering
\begin{tabular*}{\textwidth}{@{}p{1.6cm}p{5.2cm}p{5.2cm}@{}}
\toprule 
\textbf{Quality Measure} & \textbf{Domain-General Suggestions}  & \textbf{Recipe-Specific Suggestions}\\
 \midrule 
  \multirow{6}{*}{\textit{Structure}} & $r_1$: use appropriate sections and headings. & $s_1$ write ingredients and steps in separate sections.\\
  & $r_2$: enumerate all steps of the procedure. & $s_2$: enumerate the recipe steps.\\
  & $r_3$: provide all necessary materials (completeness). & $s_3$: write down all ingredients.\\
  & $r_4$: provide the materials as a list. & $s_4$: list each ingredient separately.\\
  \midrule
    \multirow{4}{*}{\textit{Clarity}} & $r_5$: give reasons for doing something. & $s_5$: give reasons for doing something.\\
  & $r_6$: describe until when to do an action. & $s_6$: describe until when to do an action.\\
  & $r_7$: describe how the result of an action looks like. & $s_6$: describe how a mixture should look or feel at a certain stage.\\
  \midrule
 \multirow{9}{*}{\textit{Specificity}} & $r_8$: specify important details regarding the materials &  $s_{8.1}$: specify whether the meat, poultry, or seafood is boned,
skinned, or prepared otherwise.\\
&& $s_{8.2}$: specify whether cheese should be shredded, crumbled, or cubed.\\
  & $r_9$: specify important details regarding the steps &  $s_{9.1}$: indicate the intensity of the heat. \\
  && $s_{9.2}$: indicate whether cookware should be covered.\\
 \bottomrule
\end{tabular*}
    \caption{Overview of suggestions used to enrich the examples; categorized into the three main aspects for quality derived from \citep{wieringa1991procedure, sato2006, Traga_Philippakos_2019, Adoniou_2013}. For the domain-specific suggestions derived from \citep{handbook}, a limited number of examples is provided (there are $38$ recipe-specific suggestions in total).}
    \label{tab:annotations}
\end{table}


We transform the suggestions into explicit rules to be able to annotate each recipe for each of the $45$ suggestions. Specifically, we classify each of the $45$ suggestions as "followed", "missing", or "not relevant" for each recipe. For example, if the recipe does not require a pan, the suggestion to ``\textit{indicate the size and type of the pan}'' is not relevant; on the other hand, if the recipe requires a ``pan'', but the size (small, medium, large) or type (frying, skillet, non-stick, ceramic, etc) are not specified, the suggestion is ``missing''. To facilitate this classification, we employ a rule-based system using regular expressions. This method allows for an automated annotation of the recipes. Our classification algorithm operates in two stages. Initially, it scans the recipe for keywords related to each suggestion (main keyword). Following the example, it would look for ``pan'' or synonyms. Subsequently, when a keyword is identified, the algorithm examines a 20-character range surrounding it to detect any mention of the specific characteristics detailed in the suggestion (supporting keywords). In our example, it would look for the size or type of the pan. This process is repeated for all suggestions, and the results are compiled into a dictionary. This dictionary reflects the status of each suggestion (followed, missing, or not relevant) for every recipe, including the specific locations where these criteria are met.

\minor{The previously described classification algorithm aims at ascertaining the presence of the specified keywords (supporting keywords) in proximity to another predetermined keyword (main keyword). We define ``proximity'' as a 20-character range to account for intervening descriptors (such as adjectives or qualifiers) that are typically positioned close to their corresponding nouns that might not be related to the suggestion. For example, for the suggestion about specifying the form of nuts (e.g., whole, halved, chopped, etc) in proximity to a nut's name (e.g., walnut, almonds), a phrase like ``slivered (form) blanched almonds (nut)'' exemplifies a case where looking at the preceding or succeeding word fails to recognize the relationship due to the intervening descriptors.  Given that the average word length in English is 4.8 \footnote{https://norvig.com/mayzner.html}, we chose a 20-character range that is approximately $4$ words apart. Empirical trials confirmed that this range effectively captures the necessary details in the majority of cases, striking a good balance between capturing essential information and 
excluding unrelated text that might pertain to other ingredients or elements rather than describing the main keyword.}

To assess the rule-based annotation performance, we conducted an evaluation using a random sample of recipe segments. Two annotators, who are also authors of this work, were involved in this process.  \minor{One of the annotators had participated in the generation of the rule-based regular expressions, while the other annotator was unfamiliar with the generation process.} The choice of annotators was a pragmatic decision that allowed us to evaluate the rule-based model without the need for recruitment of external annotators.
Per each suggestion, we randomly selected five recipe segments where the two-step annotation algorithm indicated that the suggestion was present and five segments where it was missing. The segments were shuffled and manually labeled to indicate whether the rule was being fulfilled or not.  The Cohen's Kappa score between the annotators was $0.85$ (near perfect agreement~\citep{landis_kappa}) and the average accuracy was $0.95$. \minor{We acknowledge that the choice of annotators could have induced a level of subjective interpretation. However, the random selection and shuffling of segments for annotation likely mitigated any subconscious biases. Moreover, the high inter-rater reliability indicates that the suggestions provided were clear and consistent, regardless of the annotators' prior involvement in the process.}

\subsubsection{Online Prediction and Selection}
\label{sec:onlinepipe}
The online part of the pipeline consists of retrieving a tailored comparison recipe for the user in real-time.\vspace{1mm}

\noindent \textbf{Quality Prediction}. When a participant $y$ asks for feedback on a recipe $x$, the first step consists in predicting the stars of the input recipe $S_y(x)$ using \texttt{RELEXset-Predictor}.\vspace{1mm}

\noindent \textbf{Recipe Retrieval}. In the next step, a candidate subset $SB_x$ of recipes with higher quality (i.e., a higher stars value) is retrieved from \texttt{RELEX-sugg-set}. $SB_x$ contains all the recipes with a rating withing the range $[S_y(x) + 0.4, S_y(x) + 0.8]$. For example if the rating of the input recipe $S_y(x)=1$, $SB_x$ will contain all the recipes with a standardized rating within the range $[1.4, 1.8]$. This range was chosen based on \texttt{RELEXset-Predictor}'s MAE (0.39) as we did not want $SB_x$ to contain recipes that fit within the error range of the predictor. Moreover, we wanted to show the user a peer recipe that is of better quality, but still similar enough for the user to relate to and not be discouraged by peer excellence \citep{rogers2016}. We tested the selected range in a pilot study with $10$ participants. We asked the participants to evaluate the level of the recipes seen in comparison to theirs, and the options were "much worse", "worse", "same level", "better" and "much better". None of the participants stated that the recipes were "much better", $60\%$ perceived the recipe as better, and $40\%$ as their same level.\vspace{1mm}

\noindent \textbf{Relevance Filtering}. The next stage of the pipeline consists of filtering the candidate subset $SB_x$ according to relevance. We consider that a recipe contains relevant feedback if it can exemplify how to successfully improve the input recipe $x$. To assess the relevance of the candidate recipe, we first identify the suggestions that are missing from the input recipe $x$. We then filter out from $SB_x$ the recipes that do not contain relevant feedback. We postulate that a recipe contains relevant feedback if it follows at least one suggestion that is missing from $x$. We denote as $rel\left(SB_x\right)$ the set of recipes from $SB_x$ containing relevant feedback. To exemplify the filtering stage, let us consider the following minimal example: $z$ is a recipe where the only suggestion classified as missing is "\textit{indicate the intensity of the heat}". Therefore, we will remove from $SB_z$ all recipes that do not specify the intensity of the heat when they should have. Thus, the resulting set $rel\left(SB_z\right)$ will contain only recipes that follow the suggestions: "\textit{indicate the intensity of the heat}".\vspace{1mm}

\noindent \textbf{Recipe Similarity}. The final step of the online pipeline aims to retrieve
from $rel\left(SB_x\right)$ the recipe that is most similar to $x$. We compute the recipe-recipe similarity using \texttt{BM25}~\citep{DBLP:conf/sigir/RobertsonW94}, a Bag-of-Word Information Retrieval model. Our main motivation for using BM25 instead of a LLM fine-tuned for text similarity such as \texttt{cpt-text}~\citep{DBLP:journals/corr/abs-2201-10005} is efficiency. Indeed, constraint $C_5$ enforces our pipeline to work in real-time and because in some cases $rel\left(SB_x\right)$ can contain more than $100,000$ recipes, we decided to use an efficient Bag-of-Word model. We evaluated the similarity computation time for $100$ random recipes in the worst-case scenario (with $100,000$ comparisons) and we found that the computation time was on average $0.8$ seconds ($\sigma = 0.1$ seconds) on a laptop with an Apple M1 processor. After computing the pair-wise similarities between $x$ and all recipes in $rel\left(SB_x\right)$, we return the recipe with the highest similarity.

\section{Experimental Design} 
\label{sec:study}
To evaluate \texttt{RELEX}, we conducted a controlled user study, where we asked participants to complete three procedural writing tasks in the domain of cooking recipes using our system. In the following, we will describe the study design, participants, procedure, and the employed measures in detail.
\begin{figure}[t]
    \centering
    \includegraphics[clip, trim=3.8cm 3.4cm 2.1cm 2.1cm, width=1.00\textwidth]{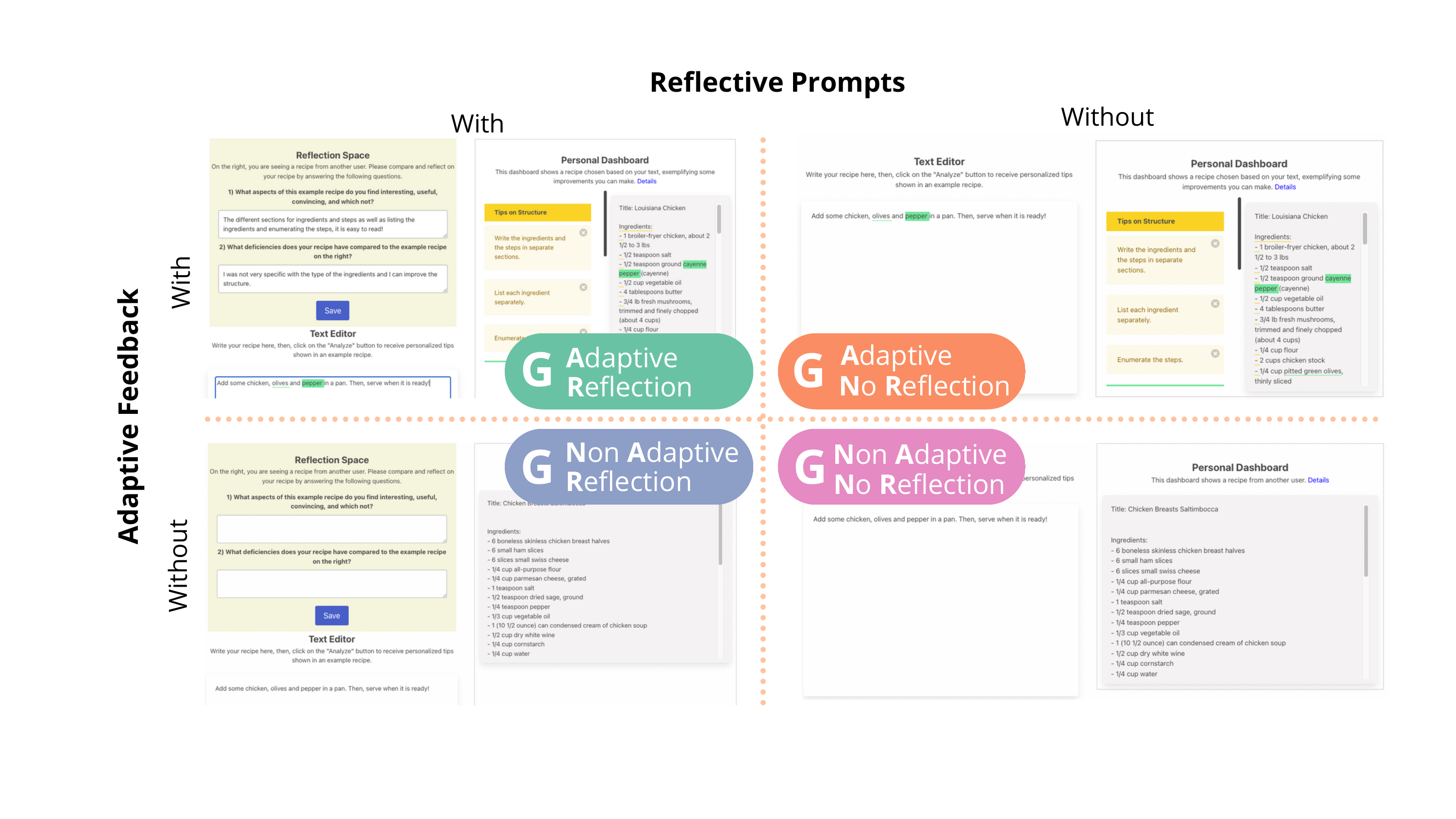}
    \caption{Illustration of the study setup using a  randomized 2 (feedback type: adaptive vs. non-adaptive) x 2 (reflection guidance: with vs. without prompts) between-subjects design resulting in four treatment groups.}
    \label{fig:four-groups}
\end{figure}

\subsection{Conditions}
\label{sec:manipulations}
We employed a fully randomized between-subjects design, encompassing two main factors: feedback type (adaptive vs. non-adaptive) and reflection guidance (with vs. without prompts). This resulted in four distinct treatment groups, each experiencing a specific combination of feedback and reflection instructions. To provide a basis for comparison, we also included a control group ($CG$), which received general static rules on how to write a cooking recipe, representing the traditional approach to support recipe writing without the provision of a peer example. 
The subjects were randomly assigned to one of the five conditions. The experiment task and questions were exactly the same for all groups; we only manipulated the adaptivity and the reflective prompts between participants. The \textbf{adaptive feedback} encompasses the tailored example recipe along with personalized in-text highlighting and explanations; and the \textbf{reflective prompts} refers to the Reflective Space where the learner is promoted to compare the recipes.


Each group used a different version of \texttt{RELEX}. Figure \ref{fig:four-groups} shows the interface for the four treatment groups \paola{experiencing varying levels of adaptive feedback and reflection guidance. The grid has two axes: Reflective Prompts and Adaptive Feedback. Each axis has two options: With and without. Thus, each quadrant represents a distinct group differentiated by the presence or absence of adaptive feedback and reflection prompts. The interface for $G_{R}^{A}$ including reflective prompts and adaptive feedback is displayed in the upper left quadrant.} $G_{R}^{A}$ used \texttt{RELEX} with all relevant functionalities including adaptive feedback (i.e., tailored example annotated with personalized in-text highlighting and explanations) and reflection prompts. The interface for $G_{NR}^{A}$ is shown in the upper right quadrant. Accordingly, $G_{NR}^{A}$ used \texttt{RELEX} without the reflection prompts. Next, as seen in the lower left quadrant, $G_{R}^{NA}$ used \texttt{RELEX} without adaptive feedback, but with reflection prompts. Lastly, $G_{NR}^{NA}$ (lower right quadrant) without reflective prompts and without adaptive feedback.  Subjects in this group were displayed a pre-selected recipe from the database. Specifically, we pre-selected five complete (in terms of structure and level of detail, see Section \ref{sec:pipeline}), but not perfect recipes (in terms of stars, see also Section \ref{sec:pipeline}) from the database. We chose to not display perfect recipes in order to keep the impression of a peer recipe. Furthermore, we made sure that the five pre-selected recipes covered a range of cooking methods (e.g., dessert, hot dish, etc.). Finally, the $CG$ did not see an example recipe; instead, an instruction manual on how to write recipes was displayed in the right pane of the tool.


\subsection{Participants}
We recruited $200$ paid participants from Prolific to conduct a controlled experiment. We chose Prolific as a platform since past research on behavioral research platforms reported the highest response quality and sample diversity for Prolific~\citep{peer2017beyond}. To avoid an overlarge diversity in our sample, we recruited participants in the age range from $18-30$ with at least an undergraduate degree as the highest completed education level. We excluded participants who did not complete the post-test or had technical problems, remaining with $187$ participants for our analyses. Table \ref{tab:participants} summarizes the demographic information per group. We did not find significant differences between the groups in terms of age (${\chi}^2(4) = 1.07, p = .90$) or gender (${\chi}^2(8) = 7.49, p = .48$) as indicated by a non-parametric Kruskal–Wallis test\footnote{We checked for normality using a Shapiro-Wilk test and verified equal variances using Levene’s test and found that for both age and gender, the assumptions of ANOVA were not satisfied.}. The median time spent on the study was $70$ minutes. Participants were paid $9$£ per hour.  

\begin{table}[t]
\centering
\resizebox{\columnwidth}{!}{\begin{tabular}{ccccccccccc}
\toprule
\multicolumn{4}{c}{\textbf{Group}}  && \multicolumn{2}{c}{\textbf{Age}} && \multicolumn{3}{c}{\textbf{Gender(\%)}}  \\\cmidrule(l){1-4} \cmidrule(l){6-7}\cmidrule(l){9-11}
\textit{ID}\textit{}  & \textit{Feedback} & \textit{Prompts} & \textit{ Users} & & \textit{Mean} & \textit{SD }      &  &  \textit{Female} & \textit{Male} &  \textit{Other} \\ 
\midrule
\vspace{2mm}
$G_{R}^{A}$ & Adaptive & Yex              & 40           && 26.7        & 2.9     &   & 57       & 40     & 3       \\
\vspace{2mm}
$G_{NR}^{A}$ & Adaptive     & No          & 35           & &26.8        & 3.3     &   & 43       & 51     & 6       \\
\vspace{2mm}
$G_{R}^{NA}$ & Non-adaptive & Yes         & 33           & &26.3        & 3.0    &    & 36       & 64     & 0       \\
\vspace{2mm}
$G_{NR}^{NA}$ & Non-adaptive & No          & 39           & &26.3        & 2.8    &    & 51       & 46     & 3       \\
\vspace{2mm}
$CG$  & None          & No          & 40           & &26.4        & 3.0     &   & 57       & 40     & 3 \\
\bottomrule
\end{tabular}
}

\caption{Demographics of participants per group (adaptive vs. non-adaptive \textit{feedback} and reflective \textit{prompts} vs. none). There are no significant differences between groups in terms of age or gender.}
\label{tab:participants}
\end{table}


\subsection {Procedure}
The experiment consisted of three main parts that were the same for all groups: (1) a pre-survey (including a pre-test), (2) three procedural writing tasks (in the domain of cooking recipes), and (3) a post-survey (including a post-test). \paola{Different from the three main tasks centered on composing cooking recipes, the pre-test and post-test were situated in a distinct domain: furniture assembly. The different domain was chosen in order to study whether the users could transfer the acquired procedural writing skills to another task}.

\vspace{1mm}
\noindent \textbf{Pre-Survey}. The experiment began with a pre-survey, where we a) controlled the effectiveness of the randomization using two different constructs (see Table \ref{tab:constructs}) and b) conducted a pre-test for procedural writing skills in the domain of furniture assembly. We started by asking each participant three questions about their previous cooking experience and documenting their recipes. Next, we captured participants' attitudes towards technology~\citep{agarwal2000time}. Both constructs were measured on a $7$-point Likert scale ($1$: totally disagree to $7$: totally agree, with $4$ as a neutral statement). Finally, we assessed participants' procedural writing skills in a transfer domain. Specifically, we asked participants to write the instructions to assemble an IKEA piece of furniture (a TINGBY table) based on a step-to-step diagram (illustration only, no text available). Participants were requested to spend five minutes solving the task.

\vspace{1mm}
\noindent \textbf{Procedural writing assignment}. In the procedural writing part of the experiments, we asked the participants to perform three cooking recipe writing tasks. The task was: \textit{"It's a Sunday afternoon and your best friend calls you with a very hectic voice: they need to prepare a dish for their family who is going to visit in the evening. Your friend asks you to provide them with three different cooking recipes to choose from. Be aware that your friend has very little cooking experience and therefore you have to write the recipe as structured and understandable as possible."}  All groups were asked to watch an introduction video on the usage of the tool before the first recipe-writing task.

\vspace{1mm}
\noindent \textbf{Post-Survey.} The experiment ended with the post-survey. First, we conducted the post-test, where participants were asked to write instructions on how to assemble a different piece of IKEA furniture (an EKET cube) based on a step-by-step diagram (illustration only). We made sure that the difficulty of assembly was similar for both tests. As in the pre-test, participants were asked to spend five minutes on the task. Next, we measured the users' perception using different constructs from literature (see Table \ref{tab:constructs}). Again, all behavioral constructs were measured on a $7$-point Likert scale ($1$: totally disagree to $7$: totally agree, with $4$ a neutral statement). Finally, participants answered five qualitative questions about the usage of the tool, the impact of \texttt{RELEX} on participants' recipe writing, and user experience.


\begin{table}[t]
\centering
\begin{tabular}{@{}p{1.5cm}p{3.0cm}p{6.5cm}@{}}
\toprule
\textbf{Time Point} &
\textbf{Construct} & \textbf{Items} \\
\midrule
\multirow{11}{*}{\textit{Pre-Survey}} & \multirow{3}{*}{Previous experience} & I am interested in cooking. \\
 &
   &
  I have experience writing recipes. \\
 &
   &
  I write recipes frequently. \\ \cmidrule{2-3}
 &
  \multirow{8}{3.0cm}{Attitude towards technology \citep{slade2020students}} &
  I like to experiment with new tools and technologies. \\
 &
   &
  If I heard about a new tool, I would look for ways to experiment with it. \\
 &
   &
  In general, I am hesitant to try out new technological tools (reversed-coded). \\
 &
   &
  Among my peers, I am usually the first one to try out new technologies and tools. \\
\midrule
\multirow{26}{*}{\textit{Post-Survey}} &
  \multirow{6}{3.0cm}{Perceived usefulness \citep{pu}} &
  By using RELEX, I can improve my procedural writing skills. \\
 &
   &
  With RELEX, I learned more efficiently aspects of procedural texts than I would have learned without it. \\
 &
   &
  I find RELEX useful. \\ \cmidrule{2-3}
 &
  \multirow{3}{3.0cm}{Affect towards use \citep{cu}} &
  RELEX makes writing recipes more interesting. \\
 &
   &
  Working with RELEX is fun. \\
 &
   &
  I like using RELEX. \\ \cmidrule{2-3}
 &
  \multirow{5}{3.0cm}{Perceived ease of use \citep{Venkatesh2008TechnologyInterventions}} &
  Learning to use RELEX takes up too much of my time (reverse-coded). \\
 &
   &
  RELEX is easy and intuitive to use. \\
 &
   &
  I found RELEX hard to understand (reverse-coded). \\ \cmidrule{2-3}
 &
  \multirow{6}{3.0cm}{Behavioral intention \citep{Venkatesh2008TechnologyInterventions}} &
  Imagining I would write a recipe in the future, I would use RELEX for it. \\
 &
   &
  Imagining I would write a recipe in the future, I would plan to use RELEX. \\
 &
   &
  Imagining a friend would write a recipe in the future, I would recommend RELEX. \\ \cmidrule{2-3}
 &
  \multirow{6}{3.0cm}{Learning gains \citep{Venkatesh2008TechnologyInterventions}} &
  My skills to write procedural texts improved after using RELEX. \\
 &
   &
  I can write better procedural recipes after using RELEX. \\
 &
   &
  RELEX helped me to learn how to write procedural recipes.\\
  \bottomrule
\end{tabular}%
    \caption{Overview of the constructs used in the pre and post-survey including detailed items.}
    \label{tab:constructs}
\end{table}

\subsection{Measures and Analysis}
\label{sec:measures}
To investigate the impact of our system, we studied \paola{learners'} writing performance on the task and the transfer task. Moreover, the impact on learners' perception was assessed using a post-survey with qualitative questions. Finally, we assessed the impact on learners' revision behavior using a keystroke analysis. 

\vspace{1mm}
\noindent \textbf{Task Performance}. To assess participants' progress in recipe writing skills, we used each participant's first recipe (i.e., their first submission) as an initial evaluation and their last revised recipe (i.e., their last submission) as a final evaluation. Specifically, we computed two scores for each recipe: the predicted stars ($S_y(x)$) and the quality score ($Q_y(x)$). The first score, the predicted stars ($S_y(x)$), was obtained using the model fine-tuned to predict the recipes' stars (\texttt{RELEXset-Predictor}, see Section \ref{sec:offlinepipe}). We gave as input the recipe written by the participant and the model returned the predicted stars. The second score is a quality/completeness score based on the {quality criteria (structure, clarity, specificity)} implemented by the set of suggestions derived in Section \ref{sec:offlinepipe}. We computed the quality score $Q_y(x)$ for a recipe $x$ from a participant $y$ based on $A_{x}$, the set of suggestions relevant to recipe $x$. For each suggestion $r_i \in A_{x}$, we computed a score $s_{y,r_i} \in \{0,1\}$, where $1$ indicates that the suggestion was followed and $0$ indicates that the suggestion was missing. We then computed the quality score as $Q_y(x)=\sum s_{y,r_i}/|A_x|$. The quality score, therefore, measures the ratio of followed rules for a recipe. \vspace{1mm}

\noindent \textbf{Transfer Performance}. To evaluate the pre-and post-test tasks, we assessed the learning objectives of procedural texts. We thus adopted the subset of suggestions regarding structure, clarity, and specificity described in Table \ref{tab:annotations}. We made adjustments to $r_8$ and $r_9$ to better suit the context of furniture assembly. Specifically, for the specificity of materials ($r_8$), we examined the level of detail provided in describing the materials, such as explicitly naming them as wood or metal. For the specificity of steps ($r_9$), we assessed how accurately the components were referred to, including terms like screws, pegs, grooves, and knobs.
Similar to measuring task performance in terms of suggestions, for each relevant suggestion $r_i$ with, $i \in \{1,...,9\}$, we computed a quality score $s_{y,r_i} \in \{0,1\}$, where $1$ indicates that the requested suggestion is followed and $0$ indicates that the suggestion is missing. The overall transfer score of the task was then calculated as $T_y(task)=\sum s_{y,r_i}/9$.

\vspace{1mm}
\noindent \textbf{Perception}. We analyzed participants' open responses with topic modeling. We used \texttt{BERTTopic}~\citep{grootendorst2022bertopic}, a technique that incorporates the contextual information of the text by clustering embeddings generated by pre-trained transformer-based language models. We used \texttt{Sentence-BERT}~\citep{DBLP:conf/emnlp/ReimersG19} to embed the sentences in the fixed-size representation required by \texttt{BERTTopic}. More specifically, we used \texttt{all-mpnet-base-v2} checkpoint from HuggingFace’s Transformers. We split the participants' answers into sentences $v$ and clustered them to obtain the topics $z$. The topics extracted by \texttt{BERTTopic} are described in terms of the most important words and their relevance. We interpreted them and assigned names to each cluster. In a next step, we computed for each sentence $v$ the probability $p_{v,z}$ of belonging to each cluster $z$. We considered that a sentence $v$ belongs to a cluster $z$ if $p_{v,z} > 0.3$ to allow for sentences to be categorized into at most three topics. We then grouped the sentences by participant $y$ to obtain the set of topics $Z_y$ for their entire text answer. As an example, assume that the answer of a participant $y$ consisted of three sentences $v_1$, $v_2$ and $v_3$ with assigned topics: $v_1$ - topics $A$, $B$, $v_2$ - topic $B$, and $v_3$ - topics $A$, $C$, $D$. In this case, the set of topics associated with the text answer of participant $y$ is $Z_y = {A,B,C,D}$.

\vspace{1mm}
\noindent \textbf{Revision Behavior}. To study users' revision behavior, we analyzed the changes made to their recipes after receiving feedback. Based on this feedback, participants were instructed to refine their recipe. This process of analysis and improvement was not limited to a single iteration; participants could engage in multiple cycles of revision. \paola{Thus, we define a "revision" to be the set of edits (deletion, insertion, and changes) executed after receiving feedback on the recipe submission. For example, if a user requests feedback, reviews an example recipe, and subsequently makes several changes to their recipe, we consider the sequence of modifications as a single revision. If the user then proceeds to engage with the "Analyze" function once more, making additional edits to the recipe, this subsequent round of alterations is classified as a second revision.} Following previous work on revision behavior and analyzing keystrokes \citep{mouchel2023understanding, zhu_analysis_2019}, we computed the following two features: \textit{revision time} (time spent revising) and \textit{total number of revisions} (number of times recipe was edited and re-submitted).

\section{Results} 
\label{sec:results}
In this study, we sought to examine the effects of adaptive feedback and reflective prompts on learners' perception (RQ1), procedural writing skills (RQ2), and revision behavior (RQ3). To achieve this, we conducted a comprehensive analysis, both quantitative and qualitative, on the data gathered from the post-survey, procedural writing assessments, and the pre- and post-test.\minor{In the following analyses we present the \textit{p}-values resulting from the analysis, the effect sizes are available at: \url{https://github.com/epfl-ml4ed/relex/tree/main/docs/effect-sizes.pdf}}. In a first preparatory step, we verified the randomization by checking for differences between the five groups at the beginning of the study. A Kruskal-Wallis test\footnote{\label{annovaassumptions}We checked for normality using a Shapiro-Wilk test and verified equal variances using Levene’s test and found the assumptions of ANOVA were not satisfied.} confirmed that there were no differences in participants' procedural writing skills as measured by their quality scores $T_y(pre)$ (see Section \ref{sec:measures}) achieved on the pretest task (${\chi}^2(4) = 4.85$, $p = .30$). For the pre-survey, we obtained the construct score by averaging the items in each construct (all factor loadings were greater than $0.7$) and found no significant differences either in participants' previous experience with documenting cooking recipes (${\chi}^2(4) = 4.83$, $p = .30$) and attitudes towards technology (${\chi}^2(4) = 4.2$, $p =.37$).
Lastly, we analyzed how long participants took to complete the study. On average, participants took $73$ minutes. Again, we found no significant differences (${\chi}^2(4) = 8.15$, $p = .09$) between the average duration time per group ($78$ minutes for $G_{R}^{A}$; $73$ minutes for $G_{NR}^{A}$; $64$ minutes for $G_{R}^{NA}$;
$79$ minutes for $G_{NR}^{NA}$; and $70$ minutes for $CG$.) 


\subsection{RQ1: Impact on Learners' Experience}
\label{sec:res-rq1}
To answer our first research question, we analyzed participants' user experience and perception. Based on the findings of~\citet{SCHWORM2006426}, we hypothesized that the perceived learning gain, usefulness, behavioral intention, and attitude towards use would be higher in the groups with adaptive feedback: $G_{R}^{A}$, $G_{NR}^{A}$ \textbf{(H1-1)}. In addition, in line with \citet{Venkatesh2008TechnologyInterventions}, we hypothesized that the perceived ease of use would be the highest in the $CG$ and the lowest in $G_{R}^{A}$ given that the $CG$ used the version with the simplest interface and functionality \textbf{(H1-2)}.

\vspace{1mm}
\noindent \textbf{Quantitative Analysis}.
In a first analysis, we compared the post-survey constructs between groups using the Kruskal-Wallis test\footref{annovaassumptions}. The results confirmed significant differences between groups concerning perceived usefulness (${\chi}^2(4) = 14.30$, $p < .01$) and behavioral intention (${\chi}^2(4) = 14.20$, $p < .01$). To further investigate the specific differences within these constructs, we performed a pairwise comparison using the Wilcoxon Rank Sum test, correcting for multiple comparisons via a Benjamini-Hochberg (BH) procedure. 

\begin{figure}[t]
    \centering
    \includegraphics[clip, trim=10cm 9.2cm 10.1cm 10.3cm, width=1.00\textwidth]{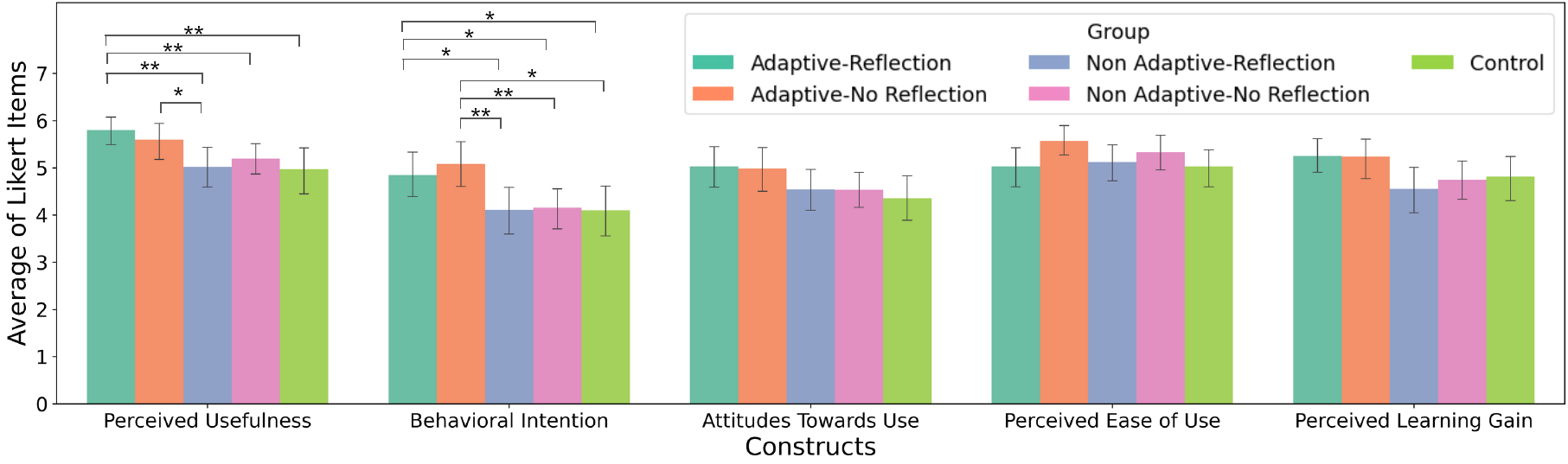}
    \caption{Post-survey answers comparison between control and treatment groups. Statistically significant differences between groups are indicated with * ($p < .05$) and ** ($p < .01$).}
    \label{fig:results_construts}
\end{figure}

Figure \ref{fig:results_construts} depicts the distribution per group and construct, with statistically significant differences marked with * ($p < .05$) and ** ($p < .01$). We observe that participants from the group receiving both adaptive feedback and reflective prompts ($G_{R}^{A}$) perceived the tool as more useful than the participants from the groups without adaptive feedback ($G_{R}^{NA}$ and $G_{NR}^{NA}$) and the control group ($CG$). Likewise, participants in $G_{NR}^{A}$ (adaptive feedback, no reflective prompts) also reported higher perceived usefulness than participants in $G_{R}^{NA}$. It is worth mentioning that the only variant between these two groups was the presence of adaptive feedback. Moreover, regarding the behavioral intention, both $G_{R}^{A}$ and $G_{NR}^{A}$ (the groups with adaptive feedback) exhibit significantly higher scores than all other groups.


In a subsequent analysis, we investigated the differences between the groups that received adaptive feedback ($G_{R}^{A}$ and $G_{NR}^{A}$) and the ones that did not ($G_{R}^{NA}$ and $G_{NR}^{NA}$). We found that the groups with adaptive feedback had significantly higher scores in four out of five constructs: perceived usefulness (${\chi}^2(1) = 11.46$, $p < .001$); attitudes toward use (${\chi}^2(1) = 5.2$, $p < .01$); behavioral intention (${\chi}^2(1) = 12.08$, $p < .001$); and perceived learning gain (${\chi}^2(1) = 6.07$, $p < 0.01$).  Interestingly, there were no significant differences in the perceived ease of use.

\vspace{1mm}
\noindent \textbf{Perception Analysis}. In our subsequent analysis, we delved into participants' open-text responses to gain deeper insights into the observed effects from the post-survey.  Specifically, we first examined the responses to the question "What did you like?". \paola{The responses reflected a positive reception of the system's features, including comparative viewing of recipes, in-text highlighting, ease of use, helpful suggestions, and educational insights.} The most frequently mentioned aspect, noted by 16\% of participants, was the opportunity to see other recipes. This feature was particularly appreciated for its comparative aspect, as highlighted by a participant from $G_{NR}^{NA}$: \textit{"[I liked] that I could compare my recipe with another, which makes you want to improve yours to a higher standard."} The next notable aspect was in-text highlighting, valued by 11\% of participants. A participant from $G_{R}^{A}$ described this feature as \textit{"useful to quickly identify areas, and it helps you learn and observe things you can improve quite intuitively."} \paola{Ease of use was also a significant point of appreciation. Participants described the system as \textit{"really intuitive, user-friendly"} and \textit{"clear, easy to use and methodical}." Additionally, participants praised the quality of the suggestions offered. Comments like \textit{"I liked that it gave useful suggestions that are actually valuable to a beginner"} and \textit{"It gives me tips and advice on how I can improve the wording and formatting of my recipe, so I can easily make these changes to improve the clarity and how clear my recipe is"} were common. Finally, the educational insights provided by the system were highlighted. One participant mentioned, \textit{"it allowed me to gain a better perspective on how to write instructions in a clearer and more concise manner. It helped me to focus on problem areas that I subconsciously missed because it has become ingrained into my writing style. Overall, I would say that it made me more aware of my writing foibles and allowed me to thus tackle those problems and improve."} Another added, \textit{"Despite reading a lot of recipes in the past, I do think that it very quickly guided me to writing more concise and easier to understand instructions. I like how quickly I learned using it, as well as how it leads you to figure out how to write good instructions rather than simply telling you a strict set of rules you must use."}}

Next, we examined participants' feedback on potential improvements to the tool. Not surprisingly, $12\%$ of participants in the control group ($CG$) proposed personalized content. One participant suggested, \textit{"I would change the recipe suggestions to be directly relevant for each written recipe. For example, after the first recipe, I added numbers to each step in the following two recipes, but still received the same feedback, so it became less useful."} Similarly, $16\%$ and $4\%$ of the participants in $G_{NR}^{NA}$ and $G_{R}^{NA}$, respectively, which were shown pre-selected recipes without using our pipeline, mentioned "adaptivity" as a potential area of improvement, suggesting to: \textit{"Limit the returned recipes to related dishes only."} \paola{Interestingly, some participants in $G_{R}^{A}$,
where participants received semantically similar examples, also expressed a desire for even more similar examples. One participant noted: \textit{"I would offer example recipes that have the same ingredients as the user's recipe."}. Another participant added \textit{"I may improve my recipe by adding ingredients that I did not previously add before to make it taste better."} Furthermore, practical suggestions for future tool iterations included the ability to scan handwritten recipes, eliminating the need to retype them, and the integration of real-time tips and advice during recipe composition. }

\vspace{1mm}
\noindent In summary, participants who received personalized examples ($G_{R}^{A}$ and $G_{NR}^{A}$) reported significantly higher perceived usefulness, attitudes toward use, behavioral intention, and perceived learning gain compared to the other conditions, confirming \textbf{(H1-1)}.  Interestingly, participants in the control group ($CG$), unaware of the other conditions, suggested the incorporation of adaptive feedback and content personalization, while participants in groups $G_{R}^{NA}$ and $G_{NR}^{NA}$ recommended showing more tailored and similar recipes. However, contrary to our expectations, there were no significant differences in the perceived ease of use between the conditions. As a result, we reject \textbf{(H1-2)} and conclude that the example-selection pipeline does not impose any perceivable burden or complexity on users.

\subsection{RQ2: Effect on Learners' Writing Performance}
\label{sec:rq2-res}
To answer the second research question, we analyzed learners' writing performance (quantitatively) and participants' open-text answers (qualitatively).
We analyzed the users' change in performance on the recipe task as well as on the furniture assembly task (transfer task). For the in-task performance, we hypothesized that learners who received adaptive feedback would outperform those who did not, because the highlighted elements and explanations reduce the cognitive load needed to capture the main elements~\citep{sweller94}, enabling participants to learn faster and perform better on the task \textbf{(H2-1)}. 
In contrast, for the performance on the transfer task, we hypothesized that the participants who received reflective prompts would perform better, because of the \textit{generation effect} that states that self-generated information is better retained and learned~\citep{RENKL2002529} \textbf{(H2-2)}. 

\begin{figure}[t]
    \centering
    \includegraphics[clip, trim=0.1cm 1.6cm 0.4cm 1.7cm, width=1.00\textwidth]{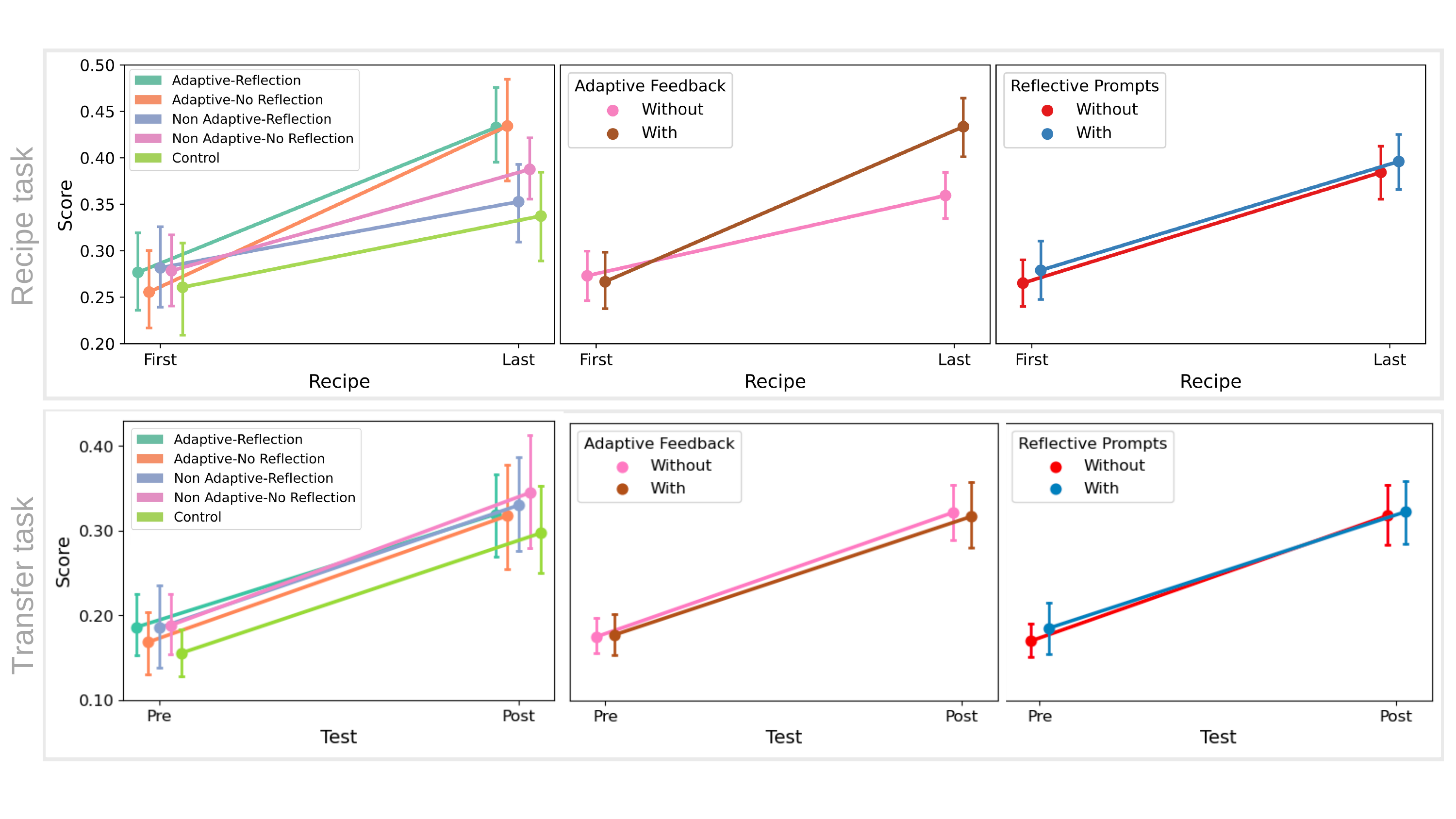}
    \caption{Performance on recipe task (in terms of quality score) and transfer task. The error bars show the standard deviation.}
    \label{fig:performance}
\end{figure}

\vspace{1mm}
\noindent \textbf{Effect on learners' task performance}. 
To test \textbf{H2-1}, we used a repeated-measures ANOVA for the predicted stars $S_y(x)$ and quality score $Q_y(x)$ (see Section \ref{sec:measures}) with the conditions ($G_{R}^{A}$, $G_{NR}^{A}$, $G_{R}^{NA}$, $G_{NR}^{NA}$ and $CG$) as the between-subjects and the test time (pre-score, post-score) as a within-subject factor. Subsequently, we proceeded with pairwise comparisons using the Wilcoxon Rank Sum test with BH corrections to investigate the differences between the various conditions.

In the quality score ($Q_y(x)$) analysis, we found a significant effect of test time ($F(1,186)=84.4, p<.0001$). \paola{Test time refers to the different measurements of the quality score through time, i.e., how the scores change from the first to the last recipe. Thus, a significant effect of test time means that the quality scores changed significantly over the course of the experiment. As seen in Figure~\ref{fig:performance} (top left), the scores in general increased from the first to the last recipes. In addition, there was also a significant interaction effect ($F(4,186)=2.6, p<.05$), which indicates that the effect of time on quality scores differed depending on the experimental condition. This is also visible in Figure~\ref{fig:performance} (top left) where some groups exhibit a steeper slope than others. This is further reinforced by a non-significant condition factor in the between-subjects analysis ($F(4,186)=1.65, p=.10$), which suggests that there were no inherent differences between the participants in the different groups. Planned pairwise comparisons confirmed the observed differences in Figure~\ref{fig:performance} (top left).} The users in $G_{R}^{A}$ improved significantly more than the users in $G_{R}^{NA}$ ($p<.05$). Likewise, users in $G_{NR}^{A}$ performed significantly better than users in $G_{R}^{NA}$ ($p<.05$) and $CG$ ($p<.05$).

In a subsequent analysis, we investigated the differences between the groups with and without adaptive feedback (Figure \ref{fig:performance} top middle) as well as with and without reflective prompts (Figure \ref{fig:performance} top right). Planned comparisons revealed that the users with adaptive feedback improved significantly more than the users without ($p<.01$) from the first to the last recipe. 

Regarding the predicted stars ($S_y(x)$) analysis, we found a significant effect of test time, with participants’ predicted stars improving significantly across recipes ($F(1,186)=19.2, p<.0001$). There was no main effect of the condition, and planned comparisons revealed no differences between the conditions.\vspace{1mm}

\textbf{Effect on learners' transfer performance}.
In a next analysis, we also used a repeated-measures ANOVA to assess performance improvements on the transfer task.  Figure \ref{fig:performance} (bottom left) illustrates the score change between participants' pre- and post-test for the five conditions. While the $CG$ seems to do worse than the other four conditions, we only found a significant effect of test time ($F(1,186)=104, p<.0001$). This suggests that on average, all the participants improved on the transfer task (see Figure \ref{fig:performance} (bottom left)). For example, $24\%$ of the participants, who did not enumerate the steps in the pre-test, enumerated the steps in the post-test. Moreover, we reviewed the tests and noted that only two participants included a title in the pre-test, while $26$ participants added it in the post-test. We also investigated the differences between the groups with and without adaptive feedback (Figure \ref{fig:performance} bottom middle) as well as with and without reflective prompts (Figure \ref{fig:performance} bottom right) and found no significant effects.


\vspace{1mm}
\noindent \textbf{Perception Analysis}.
To relate the observed effects on performance to participants' perceived performance, we again examined the survey's open-text answers.
After each recipe, participants were asked to describe the changes they made in their recipes. $20\%$ of the participants referred to enumerating: \textit{"I numbered the steps to make the order clearer. It was a good point and will allow who is cooking to quickly find the step they need"} ($G_{NR}^{A}$). Most of the consecutive popular topics referred to the recipe suggestions and explanations, for example, "specifying the size and type of pan" ($10\%$), "using more appropriate terms than add like mix, stir, beat" ($9\%$). Interestingly, despite not having direct suggestions, some participants in $G_{R}^{NA}$, $G_{NR}^{NA}$, and $CG$ made similar changes. For example, a participant in $G_{R}^{NA}$ mentioned: \textit{"I added a size measurement to my description of a baking pan because I realised it is helpful to have these details available for new bakers who are unsure of what sizes these things ought to be"}. 

In addition, as observed in the post-test, no participant in the CG mentioned adding a title and only $3\%$ of the participants in $G_{NR}^{A}$ mentioned it. In comparison, $12\%$ and $13\%$ of the participants from $G_{R}^{NA}$ and $G_{NR}^{NA}$ said they added a title; a participant from $G_{R}^{NA}$ wrote: \textit{"I [originally] did not give my recipe a title. I saw that in the example recipe and realised stating the title would help the presentation"}.

Additionally, to comprehend the impact of the reflective prompts, we examined how participants in groups $G_{R}^{A}$ and $G_{R}^{NA}$ responded regarding their utilization of these prompts. Among the participants, $27\%$ mentioned that the prompts were useful in identifying areas of improvement, with one participant expressing, \textit{"I had to actually think about where I was going wrong and what was good about the example"}. For $12\%$ of the participants, the reflective prompts acted as a means of introspection, leading them to consider ways to enhance their own recipe writing. One participant explained, \textit{"It forced me to be introspective about my own recipe writing and thus think of ways to improve my instructions."} However, a small percentage (7\%) of the participants expressed a dislike for the prompts. For instance, one participant conveyed, \textit{"Not much, the reflective questions were just a part to write what I was already thinking."} This observation could provide some insight into why we did not observe a significant effect of the reflective prompts on performance.

\vspace{1mm}
\noindent {
In summary, our findings support \textbf{H2-1} as we observed significant differences in task performance between groups with and without adaptive feedback. However, contrary to our expectations, we did not find any significant differences in task performance between groups with and without reflective prompts, leading us to reject \textbf{H2-2}. Furthermore, the results from the perception analyses indicate that participants from all groups demonstrated a good understanding of the basic elements of a procedural text.
}

\subsection{RQ3: Effect on Learners' Revision Behavior}
In addressing our final research question, we studied how users revised their recipes after receiving feedback. We formulated two hypotheses to explore this aspect. Firstly, we hypothesized that the groups with reflection prompts would invest more time in the revision process. Participants in these groups were required to answer the reflective questions, and we anticipated that this reflective practice would lead them to approach revisions with a critical mindset, spending more time contemplating potential improvements (\textbf{H3-1}). Additionally, we hypothesized that groups receiving adaptive feedback would continue revising over time, as the feedback provided would remain pertinent and applicable to their writing efforts (\textbf{H3-2}).

\vspace{1mm}
\noindent \textbf{Quantitative Analysis}.
Firstly, we investigated users' average \textit{revision time} (time spent revising the recipes). We compared the revision features between groups using the Kruskal-Wallis test\footref{annovaassumptions}, confirming that there were significant differences between groups for time (${\chi}^2(4) = 12.2$, $p < .01$). Then, to investigate the differences between groups, we performed post-hoc Wilcox pairwise comparison\footnote{correcting for multiple comparisons via BH procedure.}.

We found that users in group $G_{NR}^{NA}$ spent significantly less time revising than users in $G_{R}^{A}$ ($p < .05$) and $G_{NR}^{A}$ ($p < .05$). However, we did not find any significant difference between the groups with and without reflecting prompts, thus rejecting \textbf{H3-1}.

\begin{figure}[t]
    \centering
\includegraphics[clip, trim=0.1cm 14.1cm 0.3cm 0.4cm, width=1.00\textwidth]{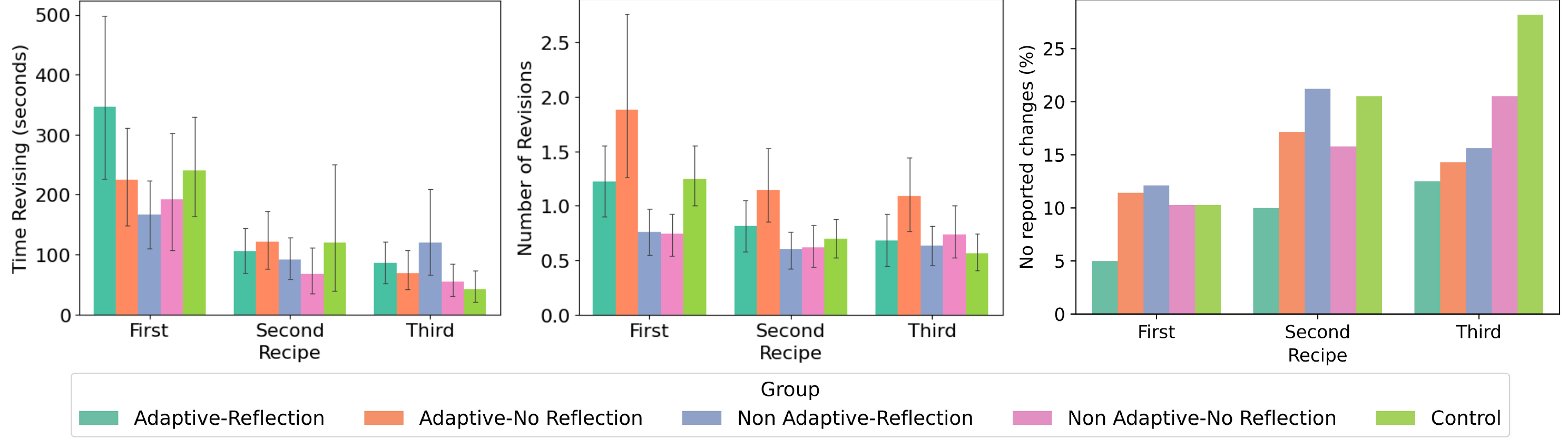}
    \caption{Revision behavior: time spent revising, number of revisions, and the percentage of declared no changes.}
    \label{fig:revision}
\end{figure}

Next, we examined how the time spent varied between the three recipes users wrote. Figure \ref{fig:revision} (left) illustrates the revision times of all five conditions for their first, second and third recipe. We observe that over time, users from all groups spend less time revising. It is worth noting that in the first recipe the users in $G_{R}^{A}$ spent on average more than twice as much time (346 seconds) as the users in $G_{R}^{NA}$ (166 seconds, $p < .05$), suggesting that the adaptive features prolongated the time users revised the recipes. 

Furthermore, when analyzing the number of revisions, we also found significant differences in the overall number of revisions per group (${\chi}^2(4) = 23.6$, $p < .001$). In particular, group $G_{NR}^{A}$ revised their recipes more than the rest of the groups ($G_{R}^{A}$, $p < .05$; $G_{R}^{NA}$, $p < .001$; $G_{NR}^{NA}$, $p < .001$; $CG$, $p < .01$). 

Moreover, we examined the number of revisions per recipe and found that there was also a general declining trend in the average number of revisions (see Figure \ref{fig:revision} (middle)). Analogously to the general results, in the first recipe, the users in $G_{NR}^{A}$ revised their recipe significantly more than the groups with no adaptive feedback ($p < .01$ for $G_{R}^{NA}$ and $G_{NR}^{NA}$). Likewise, in the second recipe, $G_{NR}^{A}$ had significantly more revisions than $G_{R}^{NA}$ ($p = .005$), $G_{NR}^{NA}$ ($p = .006$) and $CG$ ($p = .02$). Despite the fact that users in $G_{NR}^{A}$ also reduced their revision count throughout all three recipes, they consistently maintained a higher average number of revisions compared to the other groups. This finding supports hypothesis \textbf{H3-2}, indicating that certain users who received adaptive feedback still perceived it as interesting or valuable enough to ask for it again. Nonetheless, it is notable that for the first recipe, users in  $G_{NR}^{A}$ revised more than users in  $G_{R}^{A}$, despite both having adaptive feedback. It might be possible that the reflective prompts increased the cognitive load for $G_{R}^{A}$, leading to less revisions.

\vspace{1mm}
\noindent \textbf{Perception Analysis}.
As mentioned earlier, after each submission, participants were asked to describe the changes they made to improve their recipes. Figure \ref{fig:revision} (right) shows the percentage of participants reporting not making any changes for their first, second, and third recipes. We observe that for all groups, a large majority of users reported changes, with the percentage of participants not improving their recipe, increased from the first to the last recipe. Not surprisingly, group $CG$ had the steepest increase: $29\%$ of participants in this group reported making no changes to their last recipe. One participant in this group mentioned that \textit{The Analyze button just outputs the same suggestions every time, so I knew already what it wanted, and I didn't need to make any changes}. This suggests that the feedback became redundant as it was static and there were no changes. In contrast, $88\%$ ($G_{R}^{A}$) and $86\%$ ($G_{NR}^{A}$) of the participants in the adaptive feedback groups continued to report changes they made to the recipe. A big portion of the changes reported by the users (83\%), came from (or were very similar to) the suggestion given by the system. Interestingly, in the first recipe, most changes were related to the structure of the recipe, for example: \textit{"I added the ingredients list and made it step by instructions. I made these steps to make it easier to follow."}  Whereas in the second and third recipes, most comments referred to the specificity of the instructions and the steps, for example, one participant of $G_{NR}^{A}$ mentioned: \textit{"I described exactly when to move onto a next step and what to look out for in a mixture in order to proceed"}.

\vspace{1mm}
\noindent {In summary, we reject \textbf{H3-1} as we did not see the groups with reflection prompts spending more time revising. Moreover, our quantitative and qualitative analyses support \textbf{H3-2} indicating that groups with adaptive feedback perceive the example recipe and annotations as relevant, while suggestions for the other groups started to feel redundant.}

\section{Discussion and Conclusion} 
\label{sec:dicussion}
In this paper, we presented \texttt{RELEX}, an adaptive learning system for enhancing procedural writing skills. \texttt{RELEX} features a real-time retrieval pipeline, enabling personalized example-based learning at scale. Our multi-step pipeline selects higher quality and semantically relevant examples for learners based on their input
and provides suggestions on how to improve their writing. We evaluated \texttt{RELEX} with $200$ users to analyze the effects of personalized examples and reflective prompts on users' writing performance, perceived experience, and revision behavior.

\vspace{1mm}
\noindent \textbf{Impact on learners' experience (RQ1)}. Our results show that providing adaptive feedback on procedural writing skills has a positive impact on the user experience (RQ1). As we hypothesized (\paola{\textit{H1-1: Adaptive feedback will lead to heightened perceptions of learning gain, usefulness, behavioral intention, and more positive attitudes towards usage among learners}}), learners who received personalized recipes and adaptive feedback ($G_{R}^{A}$ and $G_{NR}^{A}$) judged the perceived learning gain, the perceived usefulness, the behavioral intention for continuous use, and the attitude towards use significantly better than those who did not receive adaptive feedback ($G_{R}^{NA}$ and $G_{NR}^{NA}$). These results are coherent with previous work~\citep{Wambsganss_Niklaus_Cetto_Söllner_Handschuh_Leimeister_2020}, where the group with adaptive feedback had a significantly higher intention to use. Moreover, our analysis of open answers exemplifies the positive reactions participants had towards seeing another recipe, in-text highlighted elements, and adaptive suggestions. A positive perception plays an important role in the long-term success of learning tools and their potential to foster learning~\citep{Kirkpatrick_1994}. 

Against our expectations and different from \citet{Fan_Luo_Menekse_Litman_Wang_2017}, we did not find any significant differences between the groups regarding the perceived ease of use (\paola{\textit{H1-2: The ease of use will be perceived as most favorable in the groups with simpler interfaces}}). We originally hypothesized that the users would find the complete interface (including the personalized example, adaptive explanations, in-text highlighting, and reflective prompts) hard to understand. \citet{Venkatesh2008TechnologyInterventions} define perceived ease of use as the degree to which a person believes that using the tool with be free of effort. Thus, we expected the extra features like reflection and suggestions to represent an effort for the users. Nevertheless, when analyzing the qualitative comments, the third highest-ranked topic was the "intuitiveness" of the tool. This suggests that the design iterations with users contributed to an intuitive design, where the special features and elements do not hinder the ease of use.

\vspace{1mm}
\noindent \textbf{Impact on learners' writing performance (RQ2)}. Moreover, we investigated the effects of the design elements (personalized example, adaptive feedback and prompts) on performance (RQ2). Our results confirm our hypothesis (\paola{\textit{H2-1: Adaptive feedback will improve in-task writing performance}}), showing that participants in the adaptive feedback groups improved their recipe quality and completeness significantly more than the participants in the non-adaptive groups. The perception analysis suggests that the in-text highlighted elements helped identify the areas of opportunity quickly. Previous work~\citep{VANGOG2008211} found that extra information and explanations were beneficial in terms of learning gains at first, but hindered performance later on as the information quickly became redundant. In our study, we overcome that challenge in $G_{R}^{A}$ and $G_{NR}^{A}$ by only showing explanations that are relevant based on the user's recipe. This adaptivity could also explain the observed performance differences given that $G_{R}^{NA}$-$CG$ received redundant explanations regardless of the user's input. This is in line with the perception analysis, where the participants in the $CG$ mentioned that the suggestions became less useful when they were redundant. 

We also studied whether participants in the groups with reflective prompts were able to generalize better when asked to transfer the skills to another domain (\paola{\textit{H2-2: Reflective prompts will improve the writing performance in a transfer task}}). Our results reject our hypothesis. We hypothesize that the duration of the user study was too short (only three recipes) to unfold the \textit{self-explanation effect} \citep{wong2002effects}. Alternatively, as noted by one of the participants, it is possible that even without writing, the participants were already explaining the example to themselves. 


Surprisingly, all groups improved on the transfer task (furniture assembly). We observed that, on average, participants improved their text $15\%$ in terms of quality (structure and specificity). This suggests that participants were able to grasp the principles of the learning domain (procedural writing) and apply them to a different exemplifying domain (furniture assembly). Furthermore, our results from H2-1 and the perception analysis indicate that participants also learned elements specific to the cooking domain (e.g., specifying the heat intensity). In H2-1 we observed significant differences when measuring the improvements from both content levels. \paola{We therefore hypothesize that the five approaches (experimental conditions) are similarly effective in teaching general procedural writing skills (from the learning domain). Yet, the conditions incorporating adaptive feedback also enhance learners' understanding in a specialized area within procedural writing: cooking recipe writing (exemplifying domain).}
On average, participants improved the structure and organization of their procedural text by $15\%$, including enumerating the steps, listing the materials, having separate sections for materials and steps, and adding appropriate sub-headings. According to the double-content description provided by~\citet{renkl09}, these elements belong to the content level of the learning domain of procedural writing (i.e., how to structure a procedural text in general), i.e. participants were able to grasp the fundamental structural elements of procedural writing by practicing only in the example domain.

\vspace{1mm}
\noindent \textbf{Impact on learners' revision behavior (RQ3)}. In the last analysis, against our expectations, we did not find that the use of reflective questions led to extended periods of revision. On the contrary, we found that users who received adaptive feedback spent more time revising (\paola{\textit{H3-1: Reflective prompts will increase the duration of revision times}}) than the users without adaptive feedback. Moreover, we observed that in general the time spent revising, as well as the number of revisions, decreased from the first to the last recipe. It is indeed interesting that despite the users spending less time revising, the recipes are of higher quality (as seen in \textbf{RQ2}).  As the perception analysis revealed, the users made fewer changes because they had already incorporated some of the feedback. \citet{ZHU2020103668} also observed a decline in the revision time with multiple tasks and hypothesized that the users became more familiar with the content and the feedback resulting in less time reading feedback and making changes.  

As expected, users with adaptive feedback continued to revise more in their second and third recipes (\paola{\textit{H3-2: Adaptive feedback will result in an increased number of revisions}}). The percentage of users in groups with adaptive feedback that report making no changes in the last recipe is lower than in the other groups. \citet{vangog2010} observed instructional explanations becoming redundant and irrelevant over time; it seems that providing personalized examples and annotations indeed helps reducing this effect. These results complement the results from \textbf{RQ1}, it is possible that the users perceived the tool as more useful if they engaged more with the feedback and spent more time making changes.

\vspace{1mm}
\noindent \textbf{Literature Contributions}.
 Our study contributes to and expands prior research in two main literature streams.\\ First, we contribute to the literature stream of artificial intelligence (AI) for example-based learning in heuristic domains. \tanja{Most prior research~\citep{VANGOG2008211, vangog2010, renkl09, RENKL2002529} on example-based learning uses \textit{static} examples: both the examples and explanations are created by experts and all the learners see the exact same content, independent of their input. In contrast to past literature, \texttt{RELEX} provides examples tailored to the needs of the learner in terms of topic (i.e. similar content) and skill level. Instead of providing a perfect expert example, we provide a peer example of better quality, but still attainable. Furthermore, we also personalize the instructional explanations based on the input text of the learner.} Additionally, we enhance the adaptive feedback by incorporating reflective prompts, leveraging the documented benefits found in the existing literature \citep{schworm2007learning, wong2002effects, CHI1989145, Roelle2012}.

Second, we contribute to the literature around SRL in AI systems. By including prompts for self-evaluation within the design of \texttt{RELEX}, we shed light on the combination of reflective prompts and personalized content and their effect on learning experiences and learning outcomes. Despite the qualitative comments on the helpfulness of the prompts and the positive effects from previous work~\citep{Roelle2012,schworm2007learning, vangog2010}, we did not see a significant effect on our quantitative outcome variables for perception or performance. This opens new lines of future research to investigate how to best integrate reflective prompts into adaptive systems.

\paola{\texttt{RELEX} contrasts with previous approaches to instruct procedural writing skills by focusing on personalization and adaptivity. In comparison to previous works \citep{Traga_Philippakos_2019,sato2006,alviana2019effect} where the instructional materials are static, meaning that all students received the same examples, in \texttt{RELEX} the example is chosen to cater to individual learning needs. Moreover, in comparison to instructional group approaches \citep{Traga_Philippakos_2019}, in \texttt{RELEX} each student can learn at their own pace and different from \citep{sato2006}, it does not require external readers to give feedback.
Furthermore, in contrast to other approaches of example-based learning \citep{sweller94,VANGOG2008211,renkl09,RENKL2002529}, in our work, not only do we provide a personalized example, but we also offset the common disadvantage of instructional explanations being redundant or too complex. By annotating the examples with instructional explanations adapted to the learner's prior text, we ensure their relevance.}

\vspace{1mm}
\noindent \textbf{Limitations and Future Work}. One of the big challenges of enriching examples in example-based learning is the relevance of the explanations~\citep{RENKL2002529}. Despite the participants' positive perception of the suggestions, they were extracted from "The Recipe Writer's Handbook, Revised and Expanded"~\citep{handbook} and inevitably include the authors' bias. For example, there are more suggestions for ingredients used in Western cuisine. The implication of this is that at scale, learners who write recipes from Western cuisine could benefit more from relevant suggestions. Future lines of work should investigate these biases and how to mitigate them.



\paola{Another limitation emerging from the database is that the prediction model was trained on user ratings that can be subjective. In addition, the ratings were given for a recipe as a whole, combining writing quality and taste. We examined the comments associated with the ratings and found that high-rated recipes (five stars) often had comments appreciating the clarity of instructions, as exemplified by remarks like \textit{" I really appreciate the instructions about using the spoon when cutting the potatoes.  This is a well-written recipe."}; and \textit{"This was easy enough to prepare on a worknight and assembly was so easy when following the well-written directions"}. Conversely, recipes with low ratings were often criticized for their lack of clarity and order, as indicated by comments such as \textit{"This recipe is written in a way that is impossible to attempt to follow or understand.  It is a disaster."}; and \textit{"Very frustrated with the directions. They are not orderly whatsoever."}. This suggests that even if a recipe is tasty, unclear writing can hinder its reproducibility, leading to low ratings. However, we acknowledge that a recipe with excellent writing but an unfamiliar or unappealing taste might also receive low ratings. In future studies, it would be beneficial to separate the variables taste and writing quality to more accurately assess their individual impacts on user ratings.}

\minor{This complexity extends to the predictive task, where \texttt{RELEXset-Predictor} attempts to account for both taste and writing quality, leading to only minor improvement over a static baseline.  We have therefore made our code and models publicly available\footnote{For those interested in replicating or building upon our work, we have made the implementation code and instructions for domain transfer available at \url{https://github.com/epfl-ml4ed/relex/readme.md}}, encouraging future research to enhance predictive accuracy, for example through the integration of new SOTA models. The design of the subsequent stages of the pipeline attempts to mitigate the limitations of \texttt{RELEXset-Predictor}. Overall, participants perceived the adaptive recipes as useful and edited the recipes accordingly. However more rigorous, quantitative assessments are needed to investigate the influence of the model performance and the chosen quality range on user perception.}
Furthermore, \texttt{RELEX} offers a promising approach for learners to improve their recipe writing skills by integrating both the learning domain (procedural writing) and the exemplifying domain (cooking). \paola{Despite its effectiveness, its scope is limited to these specific areas. One main takeaway for the research community is the demonstrated importance of adaptivity and personalization in example-based learning, particularly in enhancing user engagement and performance outcomes.} In future work, the example-selection pipeline can be adapted to cater to other learning domains. For instance, journal writing~\citep{Roelle2012}, high school instruction~\citep{HILBERT200854}, or argumentative writing~\citep{schworm2007learning}. \paola{Transferring \texttt{RELEX} to a different exemplifying or learning domain requires two main ingredients: 1) multiple examples with associated evaluations, ratings, or grades, and 2) domain-specific suggestions regarding example annotation. The selection pipeline (see Section \ref{sec:pipeline}) can be used to fine-tune an NLP model to predict the evaluations of the examples. Then, the model name and domain-specific suggestions can be added to the code base of \texttt{RELEX} to run the application.} By extending the tool's capabilities to various educational contexts, we anticipate a broader impact and potential benefits for learners across different domains \footnote{\label{foot:repro}For those interested in replicating or building upon our work, we have made the implementation code and instructions for domain transfer available at \minor{https://github.com/epfl-ml4ed/relex/}readme.md.}.

In the future, we envision expanding the scope and applicability of our findings by conducting replication studies in real-world settings, such as classrooms with chef apprentices. This approach would help address the ecological validity of the results and provide insights into the effectiveness of \texttt{RELEX} in practical educational contexts. Additionally, we plan to explore the long-term effects of \texttt{RELEX} by conducting a longitudinal study, assessing how repeated usage of the tool impacts learners' procedural writing skills over an extended period.

\bibliography{references_ijaied}

\section*{Statements and Declarations}

\subsection*{Funding}
This project was substantially co-financed by the Swiss State Secretariat for Education, Research and Innovation (SERI).
\subsection*{Competing Interests}

The authors have no relevant financial or non-financial interests to disclose.



\end{document}